\def\kms{km s$^{-1}$}

\def\hii{H{\sc ii}}
\def\msun{M$_\odot$}
\def\lsun{L$_\odot$}
\def\mjyb{mJy beam$^{-1}$}

\def\cmtres{cm$^{-3}$}

\def\gra{$^{\circ}$}

\def\deg{$^{\circ}$}
\def\radec{RA,Dec.(J2000)}

\def\guno{G341.220-0.213}
\def\gdos{G341.217-0.237}
\def\gtres{G341.210-0.252}
\documentclass[printer]{aa} 

\usepackage{graphicx}
\usepackage{txfonts}
\usepackage{natbib}

\begin{document}

   \title{Molecular gas and star formation towards the IR dust bubble S\,24 and its environs}

   \author{C. E. Cappa\inst{1}\inst{2},
           N. Duronea\inst{1}, 
           V. Firpo\inst{3},
           J. Vasquez\inst{1}\inst{2},
           C. H. L\'opez-Caraballo\inst{3},
           M. Rubio\inst{4}
           \and  
           M. M. Vazzano\inst{2}
           }

   \institute{\inst{1} Instituto Argentino de Radioastronom\'ia, CONICET, CCT La PLata, C.C.5, 1894, Villa Elisa, Argentina \\
           \email{ccappa@fcaglp.unlp.edu.ar}\\
           \inst{2} Facultad de Ciencias Astron\'omicas y Geof\'isicas, Universidad Nacional de la Plata, Paseo del Bosque s/n, 1900, La Plata, Argentina \\
           \inst{3} Departamento de F\'isica y Astronom\'ia, Universidad de La Serena, La Serena, Chile\\
           \inst{4} Departamento de Astronom\'ia, Universidad de Chile, Chile
           }

   \date{Received ...., 2013; accepted }

 % \abstract{}{}{}{}{} 
% 5 {} token are mandatory
 
  \abstract
   {}
 {We present a multi-wavelength analysis of the infrared dust bubble S\,24, and the extended IR sources G341.220-0.213 and G341.217-0.237 located in its environs, with the aim of investigating the characteristics of the molecular  gas and the interstellar dust linked to them, and analyzing the evolutionary state of the young stellar objects identified there and its relation to S\,24 and the  IR sources.}
   {Using the APEX telescope, we mapped the molecular emission in the CO(2-1), $^{13}$CO(2-1),  C$^{18}$O(2-1), and $^{13}$CO(3-2) lines in a region of about $5\arcmin \times 5\arcmin$ in size around the bubble. The cold dust distribution was analyzed using submillimeter continuum images from ATLASGAL and Herschel. Complementary IR and radio data at different wavelengths were used to complete the study of the interstellar medium in the region.}
   {The molecular gas distribution shows that gas linked to the S\,24 bubble, G341.220-0.213, and G341.217-0.237 has velocities between --48.0 \kms\ and --40.0 \kms, compatible with a kinematical distance of 3.7 kpc generally adopted for the region. The gas distribution reveals a shell-like molecular structure of  $\sim$0.8 pc in radius bordering the S\,24 bubble. A cold dust counterpart of the shell is  detected  in the LABOCA and Herschel-SPIRE images.  The presence of weak extended emission at 24 $\mu$m from warm dust  and  radio continuum emission projected inside the bubble indicates the existence of exciting sources and that the bubble is a compact \hii\ region. Part of the molecular gas bordering the S\,24 \hii\ region coincides with the extended infrared dust cloud SDC341.194-0.221. 
A molecular and cold dust clump is present at the interface between the S\,24 \hii\ region and G341.217-0.237, shaping the eastern border of the IR bubble. 
As regards G341.220-0.213, the presence of an arc-like molecular structure encircling the northern and eastern sections of this IR source indicates that G341.220-0.213 is  interacting with the molecular gas. The analysis of the available IR point source catalogs reveals the existence of young stellar {\bf object} (YSO) candidates linked to the IR extended sources, thus confirming {\bf their} nature as active star-forming regions.  Gas and dust masses were estimated for the different features.
The total gas mass in the region and the H$_2$ ambient density amount to 10300 \msun\ and 5900 \cmtres, indicating that \guno, \gdos, and the S\,24 \hii\ region are evolving in a high density medium. A triggering star formation scenario for the \hii\ region is investigated. 
}
   {}


\titlerunning{Cappa et al.: Multiwavelength study of S\,24}

\maketitle
%
%________________________________________________________________

\section{Introduction}%--Sect.1

\citet{churchwell06,churchwell07} identified more than 600 infrared (IR) dust bubbles in the Spitzer 8.0 $\mu$m-images of the  Galactic Legacy Infrared Mid-Plane Survey Extraordinaire (GLIMPSE, \citealt{benjamin03}) between longitudes from \hbox{--60\degr} to +60\degr. These infrared (IR) dust bubbles are bright at 8.0 $\mu$m, showing the emission of polycyclic aromatic hydrocarbons (PAHs) typical of photodissociation regions (PDRs), and enclose 24 $\mu$m emission, indicating the existence of warm dust inside \citep{watson08}. These characteristics indicate the presence of excitation sources (O- and/or B-type stars) that dissociate the molecular gas in the border of the PDRs and heat the dust through their UV photons. The dense shells surrounding many IR dust bubbles {\bf are believed to favour} the formation of  new generations of stars through the Radiatively Driven  Implosion (RDI) or the {\it collect and collapse} (CC) processes \citep{pomares+09,deharveng+09,samal+14,deharveng+12}. Indeed, detailed studies of these bubbles have shown the presence of young stellar objects in their environs,  although in some cases triggered star formation could not be proved (see for example \citealt{alexander13, dewangan13}).

%------------------------------------Figure 1
\begin{figure*}
\centering
\includegraphics[width=17.5cm]{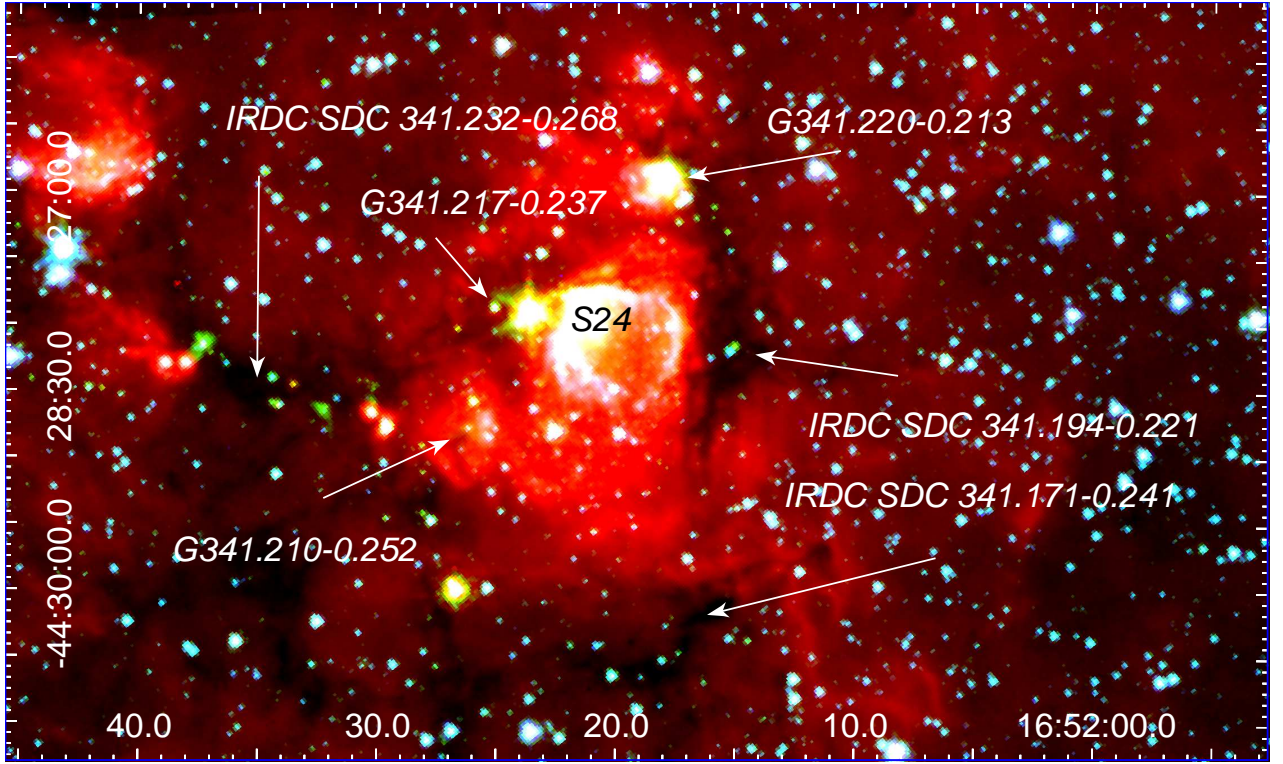}
\label{8m}
\caption{Composite image of the IR dust bubble S\,24 and its environs showing IRAC images: 3.6 $\mu$m is in blue, 4.5 $\mu$m is in green, and 8.0 $\mu$m is in red. The different features in the environment of S\,24 are indicated. }
\end{figure*}
%----------------------------------------------------

We selected the complex of IR dust bubbles S\,21-S\,24 identified by \citet{churchwell06} in the GLIMPSE images to perform a multiwavelength study of the interaction with their environs and search for young stellar objects in their vicinity. Here we report on the study of S\,24. This bubble is centered at \radec\ = (16$^h$52$^m$20$^s$, --44$\degr$28$\arcmin$10\arcsec) or {\it (l,b)} = (341\degr 203, --0\degr 23) and coincides  with the IR source IRAS\,16487-4423, which has colours typical of ultracompact \hii\ regions \citep{w&churchwell89}.

A composite image of S\,24 and its environs at 3.6, 4.5, and 8.0 $\mu$m is displayed in Fig.~\ref{8m}. At 3.6 $\mu$m, the emission arises in a faint PAH feature at 3.3 $\mu$m and in dispersed stellar light, while at 8.0 $\mu$m most of the emission originates in strong features from PAHs. The 4.5 $\mu$m band shows emission from Br$\alpha$ and Pf$\beta$, and vibrational H$_2$ lines and ro-vibrational CO lines, typical from shocked gas (see \citealt{watson08}).  At 8 $\mu$m, the bubble presents a full filamentary shell of about 28\arcsec\ in radius, brighter towards the north and east and displays a sharp eastern border. Two small and bright regions of extended emission can be identified at 8 $\mu$m in the image, named  G341.220-0.213 at \radec\ = (16$^h$52$^m$18.25$^s$, --44\degr 26\arcmin 54\arcsec), of about 10\arcsec\ in radius, and G341.217-0.237 at \radec\ = (16$^h$52$^m$23.74$^s$, --44\degr 27\arcmin 53\arcsec), of about 5\arcsec\ in radius, placed to the north and east of S\,24, respectively.  At this wavelength, the shape of G341.220-0.213 resembles a horseshoe opened toward the east. The presence of an infrared dark cloud (IRDC) identified by \citet{per&full09} as SDC341.194-0.221, is evident to the west of the bubble  where the emission at 8 $\mu$m is lacking. It is centered at \radec\ = (16$^h$52$^m$15.5$^s$, --44\degr 28\arcmin 19.4\arcsec) and extends almost 2\arcmin\ in declination. Two IRDCs were also identified by \citet{per&full09}: one at  \radec\ = (16$^h$52$^m$15.4$^s$, --44\degr 30\arcmin 6.9\arcsec) (SDC 341.171-0.241), to the south of the bubble, and the other one at  \radec\ = (16$^h$52$^m$35.7$^s$, --44\degr 28\arcmin 21.8\arcsec) (IRDC SDC 341.232-0.268). Finally, fainter emission can be detected in the whole region, encircling S\,24, G341.22-0.213, and G341.217-0.237. An almost circular  structure of about 16\arcsec\ in radius, named from hereon G341.210-0.252, is also detected in the 8 $\mu$m image at \radec\ = (16$^h$52$^m$26.5$^s$, --44\degr 28\arcmin 50\arcsec), to the southeast of S\,24. The above mentioned features are indicated in Fig.~\ref{8m}. The figure also shows that G341.220-0.213, G341.217-0.237, and part of the S\,24 bubble coincide with sources at 4.5 $\mu$m.
 
\citet{henn01} observed the region at 1.2 mm using the SEST telescope with an  angular resolution of 23\arcsec. They detected cold dust emission from G341.220-0.213 and G341.217-0.237, and an extended halo linked to  S\,24, and  estimated that the total mass in the region amounts to almost 5000 \msun\ (adopting a gas-to-dust ratio of 100 and a distance of 3.7 kpc).

Previous studies toward the region of S\,24 have shown that star formation is active. Two spots of maser emission were detected. The first spot   is a  methanol and water  maser emitter \citep{bay12,urq13,walsh14} with peak velocities between $\sim$ $-$50 \kms\ and $-$38 km s$^{-1}$ \citep{pest05,casw10,voro14,walsh14}, coincident with G341.220-0.213.  The second spot is also a methanol  maser emitter  \citep{vassilev08,bay12}   with peak velocities between $\sim$ $-$70 \kms\ and $-$16  km s$^{-1}$ \citep{l&v08,voro14}   located  at  \radec\ $\approx$ (16$^h$52$^m$15$^s$.9, $-$44\gra 28\arcmin 36\farcs 7), coincident with the IRDC SDC341.194-0.221. Although a number of EGO candidates have been identified in the S21-S24 complex \citep{cyga08}, none of them appear projected in the close environs of the IR bubble.

S\,24 is detected as a radio continuum source in the Molonglo Galactic Plane Survey at 843 MHz (MGPSJ165220-442802, \citealt{murphy07}) (synthesized beam = 65\arcsec $\times$45\arcsec) with an integrated flux density $S_{0.843}$ = 0.92$\pm$0.04 Jy.

\citet{bronf96} performed CS(2-1) line observations with an angular resolution of 50\arcsec, using the SEST telescope. They observed the molecular line in one position towards IRAS16487-4423 (\radec\ = (16$^h$52$^m$21.24$^s$, --44\degr 28\arcmin 2\farcs 5), detecting CS emission with a velocity of --43.4 \kms, indicating the presence of dense gas towards the IR source. \citet{russeil04} observed the CO(1-0) and CO(2-1) lines with an angular resolution of 42\arcsec\ and 21\arcsec, respectively, towards a position located 4\farcm 3 to the southeast of the S\,24 bubble, detecting molecular gas having velocities in the range --40 \kms\ to --37 \kms. All velocities in this paper are referred to the LSR.

Circular galactic rotation models predict that gas having velocities of --43.4 \kms\ lies at near and far kinematical distances of 3.6 kpc and 12-13 kpc, respectively (see for example \citealt{b&b93}). Similar distances can be derived from maser velocities. Following \citet{henn01} we adopt a distance of 3.7 kpc for the S\,24 region and its environment. Taking into account a velocity dispersion of 6 \kms\ for the interstellar gas, we adopt a distance uncertainty of 0.5 kpc. 

Here, we present the very first  molecular line  and dust continuum study towards S\,24 and its environs aimed at determining the distribution of the  molecular gas and the  dust linked to the bubble and their correlation between them, analyzing their physical conditions,  masses and ambient densities, and investigating the status of star formation in the region.  The presence of dense gas linked to this region makes it particularly interesting to investigate the physics and kinematics of  \hii\ regions and bubbles in dense media, allowing a better understanding of the evolution and star formation processes around these objects. Preliminary results were published in \citet{cappa13}.

\section{Data sets}%---Sect.2

\subsection{Molecular line observations}%---Sect.2.1

The molecular gas characteristics were investigated by performing $^{13}$CO(3-2) line observations (at 330.58797 GHz) of a region of 3\farcm 2 $\times$ 3\farcm 2 during  August 2009, and $^{12}$CO(2-1), $^{13}$CO(2-1), and C$^{18}$O(2-1) line observations of a region of 5$'$ $\times$ 5$'$ obtained in October 2010, with the APEX telescope using the APEX-1 and APEX-2 receivers, whose system temperatures are  150 K and 300 K, respectively. 

The  half-power beam-widths of the telescope are 30\arcsec\ (for the (2-1) lines) and 21\arcsec\ (for the (3-2) lines). The data were acquired with a FFT spectrometer, consisting of 4096 channels, with a total bandwidth of 1000 \kms\ and a velocity resolution of 0.33 \kms. The region  was observed in the position switching mode using the OTF technique with a space between dumps in the scanning direction of 9\arcsec\ for the (2-1) transition, and in the position switching mode in the (3-2) transition.  The off-source position free of  CO emission was located at \radec\ = (13$^h$33$^m$10.3$^s$, \hbox{--62\degr 2\arcmin 41\arcsec)}.

Calibration was performed using  Mars and  X-TrA sources. Pointing was done twice during observations using  X-TrA, o-Ceti and VY-CMa. The intensity calibration has an uncertainty of  10\%.

The spectra were reduced using the Continuum and Line Analysis Single-dish Software (CLASS) of the Grenoble Image and Line Data Analysis Software (GILDAS) working group\footnote{http://www.iram.fr/IRAMFR/PDB/class/class.html}. A linear baseline fitting was applied to the data. The rms noise of the profiles after baseline subtraction and calibration is 0.35 K for the (2-1) and (3-2) transitions. The observed line intensities are expressed as main-beam brightness temperatures $T_{mb}$, by dividing the antenna temperature $T_A$  by the main-beam efficiency $\eta_{mb}$, equal to 0.72 for APEX-1 and 0.82 for APEX-2 (Vassilev et al. 2008). The Astronomical Image Processing System (AIPS) package and CLASS software were used to perform the analysis.

\subsection{Dust continuum data}%---Sect.2.2

To trace cold dust emission we used far-infrared ({\sc FIR}) images from the {\em Herschel Space Observatory}\footnote{{\em Herschel} is an ESA space observatory with science instruments provided by European-led Principal Investigator consortia and with important participation from NASA} and from the APEX Telescope Large Area Survey of the Galaxy (ATLASGAL, \citealt{schull09})

\subsubsection{Herschel data}%---Sect.2.2.1

The archival data correspond to  the Hi-GAL key program (Hi-GAL:{\em Herschel} Infrared GALactic plane survey, \citealt{molinari10}, OBSIDs: 1342204094 and 1342204095).

The data have been taken in parallel mode with the instruments PACS \citep{poglitsch10} at 70 and 160\,$\mu$m, and SPIRE \citep{griffin10} at 250, 350, and 500\,$\mu$m. The chosen tile covers a field of 2.2 square degrees and is approximately centered at ({\em l,b})\,=\,(340\deg,0\deg). The angular resolutions for the 5 photometric bands spans from 8\arcsec\ to 35\farcs 2 for 70\,$\mu$m to 500\,$\mu$m. {\em Herschel} Interactive Processing Environment (HIPE v12\footnote{{\em HIPE} is a joint development by the Herschel Science Ground Segment Consortium, consisting of ESA, the NASA Herschel Science Center, and the HIFI, PACS and SPIRE consortia members, see http://herschel.esac.esa.int/HerschelPeople.shtml.}, \citealt{ott10}) was used to reduce the data, with reduction scripts from standard processing. 
To combine the two obsids (scan and cross-scan) of PACS archive data, we used the HIPE implementation script of the map merger called Scanmorphos. This script starts from Level\,1 frames and creates a map from the data. The obtained pixel sizes are 2\arcsec /px to blue and 3\arcsec /px to red. 

The zero-level of the PACS and SPIRE archive data is unknown, which may cause problems, for example to derive dust temperatures. 
The zero-level offsets of the SPIRE maps were corrected as described in \citet{bernard10}. A  proper zero-level was set bearing in mind that Planck includes the same fraction of the sky as Hi-GAL data and taking into account that the two common photometric channels of Planck and SPIRE (857 GHz\,$\sim$\,350\,$\mu$m and 545 GHz\,$\sim$\,550\,$\mu$m) are calibrated with respect to FIRAS \citep{piat02}. We used the HIPE implementation script to run destriper gain corrections and to re-calculate the absolute offset via cross-calibration with Planck-HFI data (HIPE v12, updated version by L. Conversi 2013).

To estimate fluxes, a Gaussian smoothing was applied to convolve all images at low resolution of 500\,$\mu$m (a FWHM of 35\farcs 2).

Assuming that an extended source has a spectral index $\alpha_{s}$\,=\,4 (2 for the blackbody emission plus 2 for the opacity of the dust), to convert from monochromatic intensities of point sources to monochromatic extended sources the obtained fluxes of each source were multiplied by 0.98755, 0.98741, and 0.96787 for 250, 350, and 500\,$\mu$m, respectively.   

\subsubsection{ATLASGAL data}%---Sect.2.2.2

This survey was performed at 870 $\mu$m (345 GHz) with an angular resolution of 19\farcs 2 (HPBW), using the Large Apex Bolometer Camera (LABOCA) \citep{sir09}. 
The camera operates with 295 pixels at the APEX 12-m sub-millimeter telescope. 

\subsection{Complementary data}%---Sect. 2.3

The millimeter and sub-millimeter data were complemented with Spitzer images at 3.6, 4.5, and 8.0 $\mu$m from the Galactic Legacy Infrared Mid-Plane Survey Extraordinaire (GLIMPSE) \citep{benjamin03}, and at  24 $\mu$m from the MIPS Inner Galactic Plane Survey (MIPSGAL) \citep{carey05}.
Radio continuum images at 843 MHz from the Molonglo Galactic Plane Survey (MGPS2, \citealt{murphy07}) with a synthesized beam of 65\arcsec $\times$45\arcsec, and at 1.4 GHz from the Southern Galactic Plane Survey \citep{haverkorn06} (beam size = 1\farcm 7) were used.

To investigate the existence of young stellar objects (YSO) candidates projected onto the region we analyzed infrared point sources from the Midcourse Space Experiment (MSX, \citealt{price01}), the Two Micron All Sky Survey (2MASS, \citealt{cutri03}), the Spitzer (\citealt{fazio04}), and the  Wide-field Infrared Survey Explorer (WISE, \citealt{wright10}).

%--------------------------------------------Fig. 2 
\begin{figure}
  \centering
\includegraphics[width=9cm]{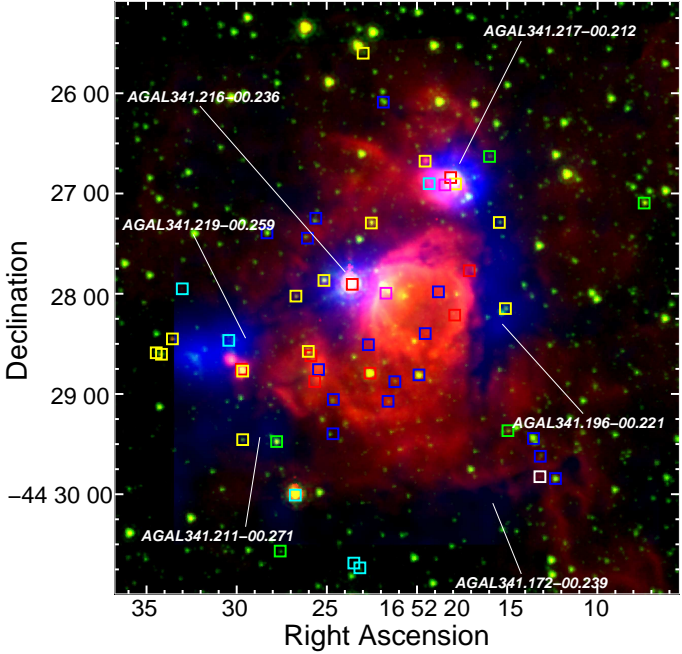}
 \caption{ Overlay of the IRAC emissions at 8 $\mu$m (in red) and 3.6 $\mu$m (in green) and  the LABOCA emission at 870 $\mu$m (in blue). Positions of identified YSOs are indicated with squares in different colours:  Class I (red) and Class II (blue) sources from the WISE database, Class I (yellow), Class II (white), and Class I/II (green) sources from the Spitzer database, 2MASS sources (cyan), and MSX sources (magenta). Dust clumps from Contreras et al. (2013) are indicated. } 
  \label{laboca}
\end{figure}
%-----------------------------------------------------

\section{YSOs projected onto the region} % Sect 3

%-------------------------------------------Table 1--------------
\begin{table*}
\centering
 {\scriptsize
\caption[]{YSO candidates projected onto S\,24 and its environs. }
\begin{tabular}{ccccccccccccc}
\hline
\hline
\multicolumn{3}{l}{\bf MSX candidates}\\
$\#$ &   [$h m s$] &  [$\circ\ \arcmin \ \arcsec$] &   {\it MSX} 
&     \multicolumn{4}{c}{Fluxes[Jy]}& & &&  \\     
 &  &  & name &  8 $\mu$m &  12 $\mu$m & 14 $\mu$m & 21 $\mu$m &  & &&Classification \\
\hline
1 & 16 52 21.7 & --44 27 59.8 & G341.2105-00.2325 &  6.2092 &  14.360 & 16.477 &  39.985  & & & & MYSO \\
2 & 16 52 18.5 & --44 26 55.0 & G341.2182-00.2136 &  2.5740 &  3.0962 & 4.1452 &  14.837  & & & & MYSO \\
\hline
\hline
\multicolumn{3}{l}{\bf 2MASS candidates:} \\
$\#$ &  $\alpha$[$h \ \ m \ \ s$]   & $\delta$[$^\circ \ \arcmin\ \arcsec$]  &    {\it 2MASS} &{\it Spitzer }  &\multicolumn{3}{c}{Fluxes[mag]} &\multicolumn{4}{c}{Fluxes[mag]} \\
 &    &    & {\it name}& {\it name}&$J$&$H$&$K_S$& $[3.6]$ & $[4.5]$ & $[5.8]$ & $[8.0]$ \\
\hline
3 & 16 52 19.32 & -44 26 54.2 & 16521932-4426541 & & 14.586 & 14.184 & 13.696 \\
4 & 16 52 23.17 & -44 30 44.1 & 16522316-4430441 & G341.1778-00.2650 & 14.653 & 14.146 & 13.631 & 12.658 & 12.258 \\
5 & 16 52 23.50 & -44 30 41.3 & 16522350-4430412 & G341.1791-00.2651 & 14.177 & 13.722 & 13.266 & 12.579 & 12.373 & 12.227 \\
6 & 16 52 26.83 & -44 30 05.6 & 16522683-4430055 & G341.1929-00.2666 & 15.733 & 15.114 & 14.377 & 13.308 \\
7 & 16 52 30.43 & -44 28 28.0 & 16523043-4428279 & G341.2207-00.2576 & 14.625 & 14.166 & 13.781 & 13.862 & 13.075 \\
8 & 16 52 33.01 & -44 27 57.1 & 16523300-4427570 & & 15.994 & 15.233 & 14.591 & 13.867 & 13.231 \\
\hline
\hline
\multicolumn{3}{l}{\bf Spitzer candidates:} \\
$\#$ &  $\alpha$[$h \ \ m \ \ s$]   & $\delta$[$^\circ \ \arcmin\ \arcsec$]  &   {\it Spitzer } &   {\it 2MASS}&\multicolumn{3}{c}{Fluxes[mag]} &\multicolumn{4}{c}{Fluxes[mag]} & Class \\
 &    &    & {\it name}& {\it name}&$J$&$H$&$K_S$& $[3.6]$ & $[4.5]$ & $[5.8]$ & $[8.0]$ \\
\hline
9  & 16 52 07.42 & -44 27 05.8 & G341.1950-00.1902 & 16520740-4427055 & 13.721 & 12.690 & 12.025 & 10.569 &  9.877 &  9.300 &  8.599 & I/II \\
10 & 16 52 13.18 & -44 29 49.6 & G341.1708-00.2324 &  &        &        &        & 13.868 & 13.610 & 10.924 &  9.793 & II \\
11 & 16 52 14.96 & -44 29 22.0 & G341.1800-00.2316 & 16521493-4429218 & 15.478 & 14.088 & $<$11.675 & 11.695 & 11.297 & 10.701 & 10.078 & I/II\\
12 & 16 52 15.10 & -44 28 08.9 & G341.1960-00.2190 &  &        &        &        & 12.845 & 10.867 &  9.955 &  9.422 & I \\
13 & 16 52 15.44 & -44 27 17.4 & G341.2077-00.2107 &  &        &        &        & 12.808 & 11.475 & 10.675 & 10.211 & I \\
14 & 16 52 15.98 & -44 26 37.9 & G341.2172-00.2050 &  &        &        &        & 12.690 & 11.839 & 11.375 & 10.424 & I/II\\
15 & 16 52 17.93 & -44 26 53.0 & G341.2176-00.2121 & 16521792-4426524 & $<$17.104 & $<$13.974 & $<$11.036 &  7.986 &  6.215 &  5.140 &  4.649 & I \\
16 & 16 52 19.55 & -44 26 40.7 & G341.2233-00.2137 &  &        &        &        & 12.436 & 12.049 &  9.074 &  7.325 & I \\
17 & 16 52 22.53 & -44 27 17.7 & G341.2210-00.2270 & 16522251-4427175 &        &        & 14.169 & 10.627 &  9.801 &  9.094 &  8.498 & I \\
18 & 16 52 22.98 & -44 25 36.1 & G341.2437-00.2101 &  &        &        &        & 12.595 & 12.343 & 11.746 &  9.833 & I \\
19 & 16 52 25.16 & -44 27 52.1 & G341.2186-00.2391 &  &        &        &        & 11.203 &  9.876 &  8.916 &  8.330 & I \\
20 & 16 52 26.03 & -44 28 34.6 & G341.2111-00.2486 &  &        &        &        & 11.410 & 10.145 &  9.104 &  8.676 & I \\
21 & 16 52 26.71 & -44 28 01.5 & G341.2195-00.2443 &  &        &        &        & 13.473 & 11.655 & 10.182 &  9.131 & I \\
22 & 16 52 27.59 & -44 30 34.1 & G341.1884-00.2733 &  &        &        &        & 12.480 & 11.695 & 11.129 & 10.675 & I/II\\
23 & 16 52 27.78 & -44 29 28.5 & G341.2028-00.2622 & 16522776-4429282 & $<$16.378 & 14.500 & 12.353 & 10.098 &  9.342 &  8.706 &  8.303 & I/II\\
24 & 16 52 29.66 & -44 28 46.3 & G341.2154-00.2590 &  &        &        &        &  9.366 &  8.211 &  7.294 &  6.387 & I \\
25 & 16 52 29.66 & -44 29 27.4 & G341.2066-00.2663 &  &        &        &        & 12.577 & 11.575 & 10.803 & 10.144 & I \\
26 & 16 52 33.55 & -44 28 27.0 & G341.2269-00.2646 &  &        &        &        & 13.792 & 11.341 &  9.639 &  8.569 & I \\
27 & 16 52 34.16 & -44 28 36.2 & G341.2261-00.2676 & 16523407-4428356 & 12.694 & 12.351 & 12.242 & 11.707 & 10.458 &  9.395 &  8.610 & I \\
28 & 16 52 34.45 & -44 28 35.3 & G341.2268-00.2681 &  &        &        &        & 12.884 & 11.260 & 10.448 & 10.137 &  I \\
\hline
\hline
\multicolumn{3}{l}{\bf WISE candidates:} \\
$\#$ &  $\alpha$[$h \ \ m \ \ s$]   & $\delta$[$^\circ \ \arcmin\ \arcsec$]  &   {\it WISE } &   {\it 2MASS}&\multicolumn{3}{c}{Fluxes[mag]} &\multicolumn{4}{c}{Fluxes[mag]} & Class\\
 &    &    & {\it name}& {\it name}&$J$&$H$&$K_S$& $[3.4]$ & $[4.6]$ & $[12.0]$ & $[22.0]$ \\
\hline
29 & 16 52 12.325 & -44 29 50.65 & J165212.32-442950.6 & 16521231-4429505 & $<$17.338 & 13.789 & 11.393 & 10.100 &  9.697 &  7.192 &  4.469 & II\\
30 & 16 52 13.168 & -44 29 37.35 & J165213.16-442937.3 & 16521312-4429365 & $<$15.988 & $<$14.795 & 13.835 & 11.981 & 11.412 &  6.822 & 3.941 & II\\
31 & 16 52 13.541 & -44 29 26.67 & J165213.54-442926.6 & 16521357-4429260 & $<$16.221 & $<$12.706 & $<$10.475 &  9.009 &  8.647 &  7.232 &  4.362  & II\\
32  & 16 52 17.106 & -44 27 46.29 & J165217.10-442746.2 & &        &        &         & 11.963 & 10.817 &  6.243 &  3.134 & I \\
33 & 16 52 17.911 & -44 28 12.87 & J165217.91-442812.8 & &        &        &        & 10.671 &  9.222 &  4.283 &  0.900 & I\\
34 & 16 52 17.961 & -44 26 52.95 & J165217.96-442652.9 & 16521792-4426524 & $<$17.104 & $<$13.974 & $<$11.036 &  8.084 &  4.888 &  2.607 &  -1.190 & I\\
35 & 16 52 18.812 & -44 27 58.92 & J165218.81-442758.9 & & & & & 7.601 & 7.286 & 2.830 & -0.197 & II \\
36 & 16 52 19.544 & -44 28 23.91 & J165219.54-442823.9 & 16522015-4428284 & 13.624 & 11.609 & 10.718 & 8.584 & 8.013 & 3.537 & 0.262 & II \\

37 & 16 52 19.890 & -44 28 48.64 & J165219.89-442848.6 & 16521984-4428486 & 14.585 & $<$11.779 & $<$10.221 &  8.933 &  8.581 &  5.674 &  1.653 & II\\
38 & 16 52 21.228 & -44 28 52.70 & J165221.22-442852.6 & 16522121-4428522 & $<$16.713 & $<$15.318 & 13.912 & 11.494 & 10.541 &  5.786 &  2.482 & II\\
39 & 16 52 21.603 & -44 29 04.46 & J165221.60-442904.4 & 16522157-4429046 & $<$15.992 & $<$14.358 & 13.513 & 11.138 & 10.539 &  5.721 &  2.730  & II\\
40 & 16 52 21.886 & -44 26 02.99 & J165221.88-442602.9 & 16522182-4426053 &  $<$17.576 & 14.702 & 12.012 & 10.136 &  9.590 &  8.005 &  4.540 & II\\
41 & 16 52 22.602 & -44 28 47.58 & J165222.60-442847.5 & 16522260-4428476 &  $<$17.062 & 14.859 & 11.133 &  8.627 & 7.384 &  5.155 &  1.507 &I \\
42 & 16 52 22.696 & -44 28 30.59 & J165222.69-442830.5 & &        &        &        &  9.250 &  8.748 &  4.270 &  0.807 & II\\
43 & 16 52 23.588 & -44 27 54.44 & J165223.58-442754.4 & 16522366-4427538 &  $<$14.662 &  $<$13.571 & 11.702 &  7.575 &  4.815 &  2.094 &  -0.978 & I\\
44 & 16 52 24.638 & -44 29 03.36 & J165224.63-442903.3 & 16522467-4429039 & 14.852 & 12.931 & 12.028 & 10.809 & 10.439 &  5.880 &  2.396  & II\\
45 & 16 52 24.659 & -44 29 23.87 & J165224.65-442923.8 & 16522482-4429238 &  $<$15.985 &  $<$15.266 &  $<$14.017 & 11.731 & 11.302 &  7.970 &  5.320 & II\\
46 & 16 52 25.457 & -44 28 45.46 & J165225.45-442845.4 & &        &        &        & 10.425 &  9.442 &  4.848 &  1.575 & II\\
47 & 16 52 25.605 & -44 27 14.89 & J165225.60-442714.8 & &        &        &        & 11.632 & 10.663 &  7.596 &  3.503 & II\\
48 & 16 52 25.668 & -44 28 52.61 & J165225.66-442852.6 & 16512576-4428529 & 16.343 &  $<$15.243 &  $<$13.988 & 10.862 &  9.794 &  5.340 &  2.220  & I\\
49 & 16 52 25.880 & -44 28 37.18 & J165225.87-442837.1 & 16522577-4428393 &  $<$15.257 &  $<$14.320 & 14.104 & 11.114 &  9.575 &  5.175 &  2.238  & I\\
50 & 16 52 26.071 & -44 27 26.80 & J165226.07-442726.8 & 16522616-4427259 &  $<$15.232 &  $<$14.120 & 13.194 & 11.546 & 10.794 &  6.597 &  3.390  & II\\
51 & 16 52 28.408 & -44 27 23.43 & J165228.40-442723.4 & 16522829-4427235 &  $<$17.943 & 14.305 & 11.736 & 10.336 &  9.777 &  8.372 &  5.300  & II \\
52 & 16 52 29.731 & -44 28 45.25 & J165229.73-442845.2 & 16522964-4428459 &  $<$14.407 &  $<$13.557 &  $<$12.079 &  9.495 &  8.082 &  4.686 &  1.089  & I\\
\hline
\end{tabular}
}
\label{ysos-todos}
\end{table*}
%--------------------------------------------------------------

As mentioned in Sect.1, maser emission linked to star-forming regions has been detected in the region. Bearing this in mind, we decided to search for  YSO candidates using the MSX, 2MASS, WISE, and Spitzer point source catalogues. To accomplish this we analyzed the IR sources in a  region of 3\arcmin\ in radius centered at \radec\ = (16$^h$52$^m$20$^s$, --44\degr 28\arcmin 00\arcsec). The results are included in Table 1 and their positions are shown in Fig.~\ref{laboca} with different symbols. 

To identify YSO candidates in the  MSX catalog we followed the criteria of \citet{lumsd02}, which takes into account their loci in the (F$_{21}$/F$_8$, F$_{14}$/F$_{12}$) diagram. Two candidates were found, classified as massive young stellar objects (MYSO), one of them linked to G341.220-0.213 and the other one to the brightest section of the S\,24 bubble. 

\begin{figure}
\includegraphics[width=10.5cm,angle=-90]{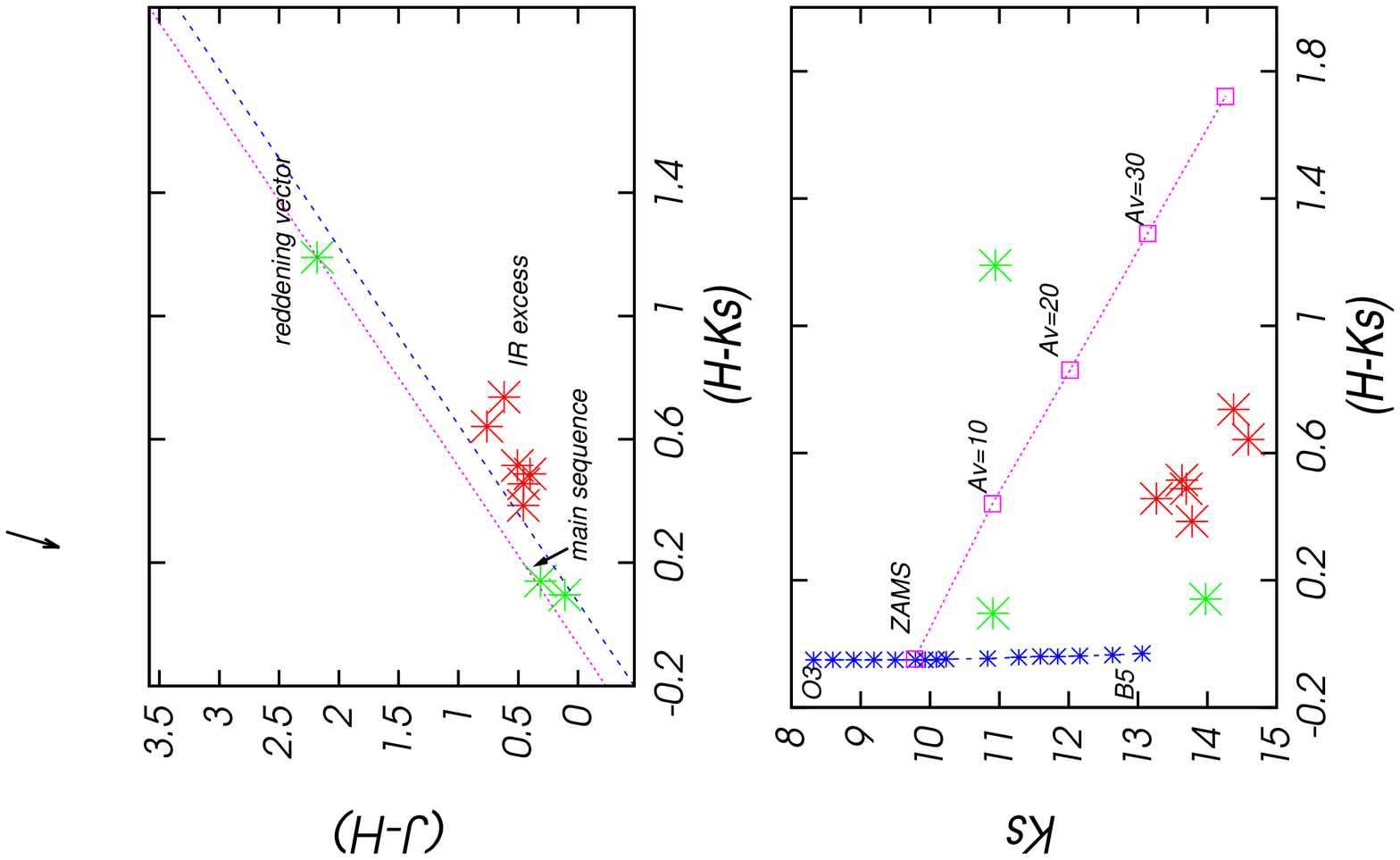}
\includegraphics[width=7cm]{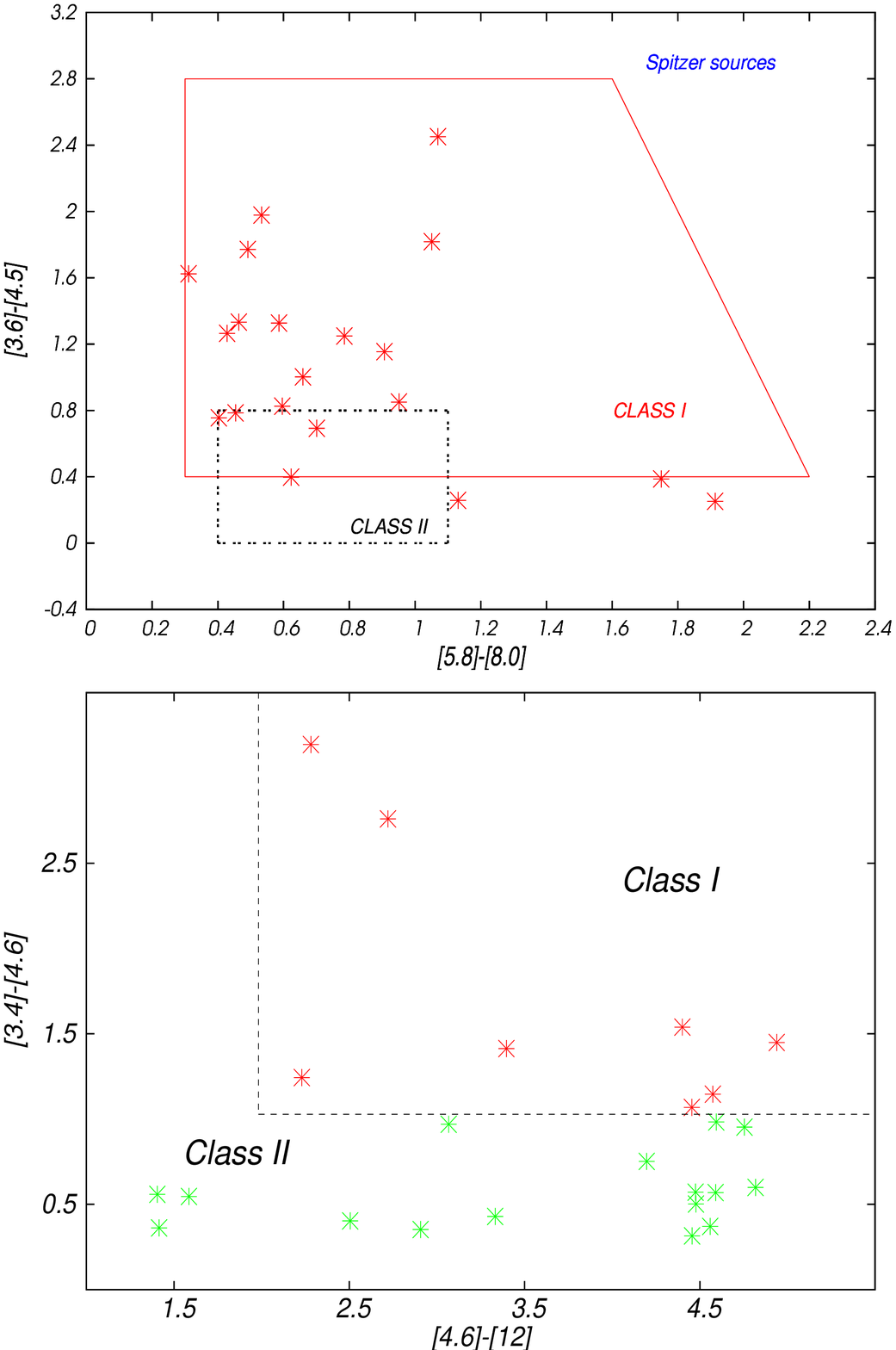}
\caption{{\it Upper panels:}  [H-K$_s$, J-H] and [H-K$_s$, K$_S$] diagrams of the 2MASS YSO candidates indicated by red crosses. Green asterisks correspond to main sequence stars projected onto the S\,24 bubble (see Sect. 7.2). The dashed lines show the reddening vectors corresponding to  M0 III and  B2 V stars in the CC-diagram, and to  B2 V stars in the CM-diagram. The squares  are placed at intervals of ten magnitudes of visual extinction. 
{\it Middle panel:} Location of YSO candidates from the Spitzer database. Crosses represent Class I and II sources. 
{\it Bottom panel:} YSO candidates from the WISE database. Class I and II sources  are indicated in red and green, respectively. }
\label{cc-diagrams}
\end{figure}

To search for YSO candidates in the 2MASS catalog, we have constructed an [H-K$_S$, J-H] diagram using only sources with photometric quality A or B. Six candidates were identified. The two upper panels of Fig.~\ref{cc-diagrams} show the 2MASS color-color (CC) and color-magnitude (CM) diagrams with the YSO candidates represented by red asterisks. 

%------------------------------------Figure 3
%----------------------------------------------------

Regarding the Spitzer database, we have performed a search for candidates adopting the criteria by \citet{allen04}, which allows to discriminate IR sources according to a Class scheme: Class I are protostars surrounded by dusty infalling envelopes and Class II are objects whose emission is dominated by accretion disks. Considering only sources detected in the four {\it Spitzer}-IRAC bands, we have investigated their evolutionary stage by analyzing their location in a color-color diagram  (middle panel of Fig. \ref{cc-diagrams}) following the color criteria determined empirically by \citet{allen04}.   Class II candidates lie in the {\bf region} \hbox{0 $\lesssim$ [3.6]-[4.5] $\lesssim$0.8} and  \hbox{0.4 $\lesssim$[5.8]-[8.0] $\lesssim$1.1}, and their IR excess might be produced by accretion disks around the stellar {\bf objects}. Sources outside the  domain of Class II objects   are identified as Class I candidate objects,  and {\bf their} IR excess might be {\bf due to} circumstellar envelopes  around the star.
 In our study we have identified 14 Class I candidates, 5 Class I/II candidates, and one Class II. The middle panel of Fig.~\ref{cc-diagrams} shows the position of the 20 Spitzer sources in the color-color [3.6]-[4.5] vs. [5.8]-[8.0] diagram. In this panel, the area delineated by thin lines corresponds to the locus of Class I objects, while the area within thick dashed lines corresponds to class II objects.

Finally, with the aim of identifying  candidate YSOs using photometric data from the {\it Wide-field Infrared Survey Explorer}  survey,  which maps the whole sky in four bands centered at 3.4, 4.6, 12, and 22 $\mu$m, we have used the criteria  of  \citet{koenig12},  using  [3.4]$-$[4.6]  vs [4.6]$-$[12] diagrams,  as follows: after removing contamination arising from background objects like galaxies (very red in [4.6]$-$[12]), broad-line active galactic nuclei (of similar colours as YSOs, but distinctly fainter) and resolved PAH emission regions (redder than the majority of YSOs),  we identified infrared excess sources demanding that
\[ \begin{array}{ll}
[3.4] - [4.6] - \sigma_1 & > 0.25\\

[4.6] - [12.0] - \sigma_2  & > 1.0

\end{array} \]
where [3.4], [4.6], and [12.0] are the WISE bands 1, 2, and 3, respectively, and  $\sigma_1 = \sigma([3.4]-[4.6])$ and $\sigma_2 = \sigma([4.6]-[12.0])$ indicate the combined errors of  [3.4]$-$[4.6] and [4.6]$-$[12.0] colors, added in quadrature.  Class I sources are a sub-sample of this defined by
\[ \begin{array}{ll}
[3.4] - [4.6] & > 1.0 \\

[4.6] - [12.0] & > 2.0

\end{array} \]
The rest of the sources are then Class II objects. Identified Class I and Class II candidates are shown in the bottom panel of Fig. \ref{cc-diagrams}.  In this panel, the separation between the location of ClassI and Class II sources is indicated by the dashed lines.  

As can be seen from Fig.~\ref{laboca}, four candidates (one MYSO, one 2MASS source, one SPITZER Class I, and one WISE Class I) appear projected onto G341.220-0.213. This fact along with the presence of maser emission towards the region confirms that this is an active star-forming region. Three candidates coincide in position with the S\,24 bubble, while only a WISE Class II candidate is projected onto G341.217-0.237. Most of the candidates appear to the east of the IRDC SDC341.193-0.221 and are projected onto  emission at 8 $\mu$m, Finally, two Spitzer candidates coincide with the IRDC SDC341.194-0.221 and other three candidates appear projected onto the borders of \gtres. 

It is clear from Table 1 that there is little correspondence among 2MASS, Spitzer, and WISE candidate YSOs (coincidences: \#24 with \#52; and \#2 with \#15 and \#34). {\bf Regarding Spitzer and WISE catalogues, in spite of the similar wavelengths of two of the bands and in the different angular resolution of the catalogues, an analysis of the coincidence of Spitzer and WISE YSOs identified in Table 1 shows that 33\% of the WISE and Spitzer sources do not have a counterpart in the Spitzer and WISE, respectively, catalogues. Additionally, 35\% of the WISE and Spitzer sources coincide with Spitzer and WISE sources, respectively,  which are undetected in at least one band and can not be properly classified. For the rest 17\%, there is coincidennce between Spitzer and WISE sources, but their counterparts are not classified as Class I or II. }

A more detailed analysis will be performed in connection to the molecular dust and cold dust in the next sections.

\section{The distribution of the interstellar dust}%---Sect.4

The distribution of the interstellar dust was investigated using data at different wavelengths in the IR. 

Figure~\ref{laboca}  shows a superposition of the  IRAC 8 $\mu$m (in red) and 3.6 $\mu$m (in green) emissions,  and the ATLASGAL 870 $\mu$m emission (in blue). Emission at 870 $\mu$m reveals the presence of cold dust. This emission approximately encircles the S24 bubble. The dust clumps identified at this wavelength by \citet{contre13} are indicated. AGAL341.216-00.236 and AGAL341.217-00.212 are bright, coincide with G341.217-0.237 and G341.220-0.213, respectively, and correlate with the emission at 1.2 mm detected by \citet{henn01}. Weaker emission at 870 $\mu$m (AGAL SDC341.196-00.221) coincides in position with the IRDC SDC341.194-0.221 placed to the west of the bubble following its sharp eastern border, and  is evident due to the absence of  8 $\mu$m emission in Fig.~\ref{laboca}. Additionally, two  cold dust clumps are present to the south and southeast of S\,24, at  \radec\ = (16$^h$52$^m$19.6$^s$, --44\degr 30\arcmin 38\farcs 5) (AGAL SDC341.172-00.239), and \radec\ = (16$^h$52$^m$31.71$^s$, --44\degr 29\arcmin 26\farcs 1) (AGAL SDC341.211-00.271). AGAL SDC341.219-00.259, located at \radec\ = (16$^h$52$^m$32$^s$, --44\degr 28\arcmin 30\arcsec), coincides with a section of the IRDC SDC341.232-0.268 and is being analyzed by Vasquez et al.(2015, to be submitted). \gtres\ does not coincide with emission at 870 $\mu$m. 

%--------------------------------------------Fig. 4
\begin{figure}
  \centering
  \includegraphics[width=9cm]{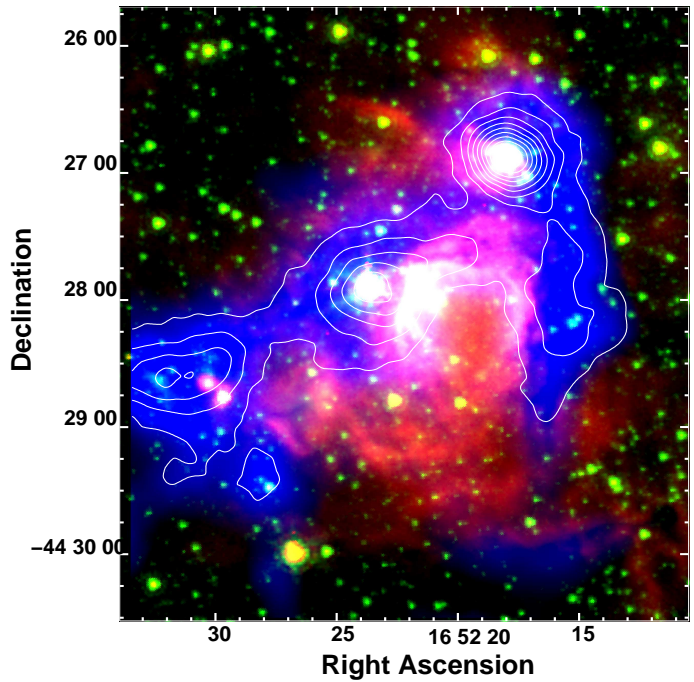}
 \caption{Composite image of the IRAC emission at 4.5 (in green) and 8.0 $\mu$m (in red), and the Herschel emission at 250 $\mu$m (in blue). Contours correspond to the emission at 870 $\mu$m and goes from 0.5 MJy sr$^{-1}$. }
  \label{herschel}
\end{figure}
%-----------------------------------------------------

Figure ~\ref{herschel} displays an overlay of the emission at 870 $\mu$m (in contours) and the emissions at 4.5 (in green) and 8.0 $\mu$m (in red) from Spitzer, and 250 $\mu$m (in blue) from Herschel. Dust clumps identified at 870 $\mu$m have their counterparts at 250 $\mu$m, although the emission at 250 $\mu$m, which also shows the distribution of cold dust, appears more extended. The brightest dust clumps at 250 $\mu$m coincide with G341.220-0.213 and G341.217-0.237, and are also detected at 350 and 500 $\mu$m (the images are not shown here). The S\,24 bubble appears clearly surrounded to the east and north by SPIRE emission.  Herschel emission at these wavelengths is also projected onto the IRDC identified to the west of S\,24, which  can be defined as a Herschel bright IRDC \citep{peretto10}. Faint extended emission at 250 $\mu$m also coincides with the IRDC SDC341.171-0.241, bordering the weak halo of 8 $\mu$m emission, and near  \radec\ = (16$^h$52$^m$28$^s$, --44\degr 29\arcmin 30\arcsec), in coincidence with a Spitzer point source identified at 4.5 $\mu$m. In summary, the correlation between 250 and 870 $\mu$m is excellent. 

\begin{figure}
  \centering
  \includegraphics[width=8cm]{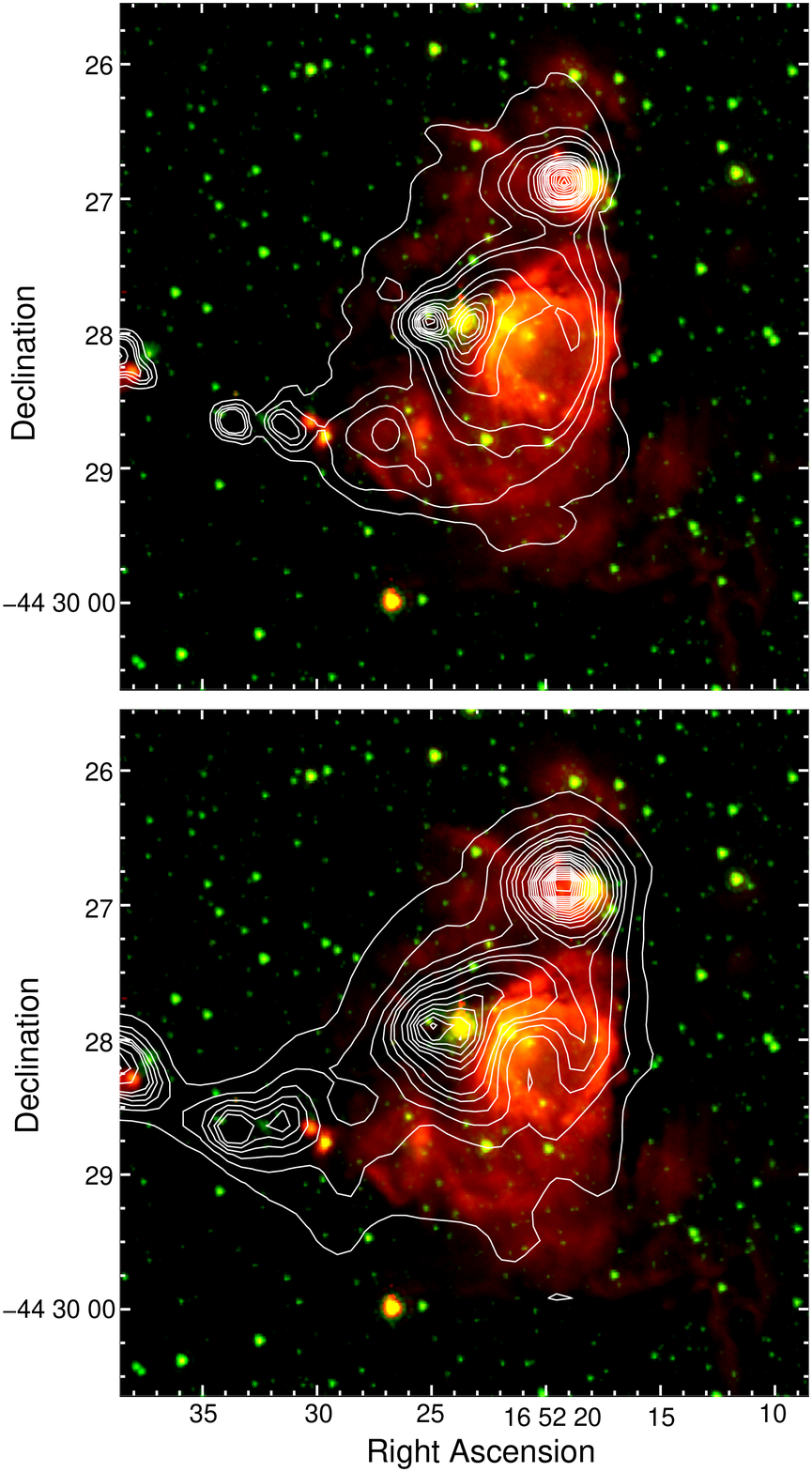}
 \caption{{\it Uper panel:} Overlay of the PACS emission at 70 $\mu$m (contours) and the IRAC 8 $\mu$m emission (colorscale). Contours are 0.3 to 0.9 Jy px$^{-1}$, in steps of 0.2  Jy px$^{-1}$; 2 to 7  Jy px$^{-1}$, in steps of 1  Jy px$^{-1}$; 9, 11, 13 to 33 Jy px$^{-1}$, in steps of 4  Jy px$^{-1}$. {\it Bottom panel:} Overlay of the PACS emission at 160 $\mu$m (contours) and the IRAC 8 $\mu$m emission (color scale). Contours are 2 to 7 Jy px$^{-1}$, in steps of 1  Jy px$^{-1}$; 10 to 25  Jy px$^{-1}$, in steps of 3  Jy px$^{-1}$; and 30 to 65 Jy px$^{-1}$, in steps of 5  Jy px$^{-1}$. }
  \label{pacs}
\end{figure}

The PACS emission at 70 and 160 $\mu$m differs significantly from the emission at larger wavelengths (Fig.~\ref{pacs}). The brightest source at these wavelengths coincides with G341.220-0.213.  G341.217-0.237 also emits at 70 and 160 $\mu$m, while a  bright source is also detected to the east of G341.217-0.237. Faint emission correlates with the S\,24 bubble, mainly at 70 $\mu$m. This emission delineates the sharp western border of the bubble.  The IRDC SDC 341.194-0.221 can not be identified either at 70 $\mu$m or at 160 $\mu$m. Emission at 70 $\mu$m is also projected onto \gtres, located at 16$^h$52$^m$26.5$^s$, --44\degr 28\arcmin 50\arcsec, but it is lacking at 160 $\mu$m and higher wavelengths. The presence of emission at 70 $\mu$m and its absence at 160 $\mu$m suggests the existence of dust at a higher temperature. 

Dust temperatures T$_{\rm dust}$ for the region using  the PACS images at 70 $\mu$m and 160 $\mu$m were estimated  by convolving the image at 70 $\mu$m down to the angular resolution at 160 $\mu$m, applying color corrections, subtracting a background local to the S\,24 bubble, and assuming that the emission is optically thin. 

The {\it color-temperature} map  was constructed as the inverse function of the ratio map of Herschel 70 $\mu$m and 160\,$\mu$m color-and-background-corrected maps, i.e., {\bf T$_{\rm dust} = f(T)^{-1}$} (see details in \citealt{preibisch12,ohlendorf13}). In the optically thin thermal dust emission regime {\bf $f(T)$} has the following parametric form:
\begin{equation}
f(T) = \frac{S_\nu({\nu=70\,\mu m})}{S_\nu({\nu=160\,\mu m})} = \frac{B_\nu(70\,\mu m,{\rm T})}{B_\nu(160\,\mu m,{\rm T})} \left( \frac{70}{160} \right) ^{\beta_{\rm d}}
\end{equation}
\noindent where $B_\nu(\nu,{\rm T})$ is the Black Body Planck function and $\beta_{d}$ corresponds to the spectral index of the thermal dust emission. The pixel-to-pixel temperature was calculated assuming $\beta_{\rm d}=2$. This is a typical value adopted for irradiated regions (see \citealt{preibisch12}).

\begin{figure}
  \centering
  \includegraphics[width=8cm]{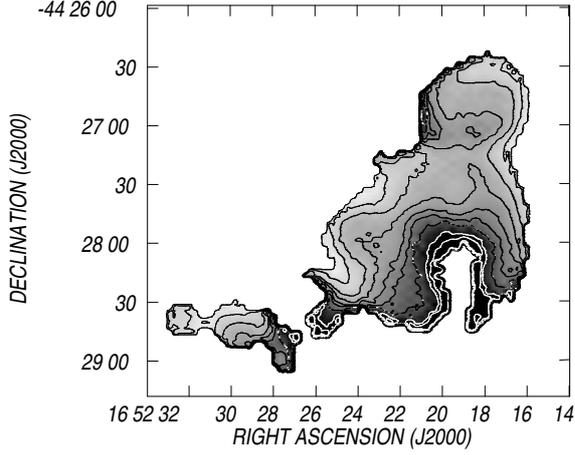}
 \caption{Dust temperature map derived from the Herschel emission at 70 and 160 $\mu$m. Grayscale goes from 20 to 50 K. Contour levels correspond to 20 to 32 K, in steps of 3 K, 35, 40, 50, and 60 K. Darkest regions indicate highest dust temperatures.}
  \label{tempera}
\end{figure}

The obtained dust temperature map is shown in Fig.~\ref{tempera}. We were able to estimate dust temperatures for G341.220-0.213, G341.217-0.237, G341.210-0.252, and the S\,24 bubble. Dust temperatures using this method were almost impossible to obtain for the IRDCs due to their low emission at 70 $\mu$m. Values for \guno\ are in the range 26-30 K, while for \gdos\, temperatures are about 29 K. As regards to the S\,24 bubble, the image reveals a clear gradient in dust temperatures (from 26 to 60 K): the higher the temperature, the closer to the center of the bubble, a fact compatible with the presence of ionizing sources linked to S\,24. Dust temperatures for S\,24 are compatible with values derived by \citet{watson10} for the interior of IR dust bubbles based on images at 24 and 70 $\mu$m from MIPS and by \citet{anderson12} for \hii\ regions using Herschel data. Dust temperatures for \gdos\ are about 28 K. 
 
%--------------------------------------------Fig. 5 
%-----------------------------------------------------

%--------------------------------------------Fig. 6 
%-----------------------------------------------------

\begin{figure}
  \centering
  \includegraphics[width=7cm]{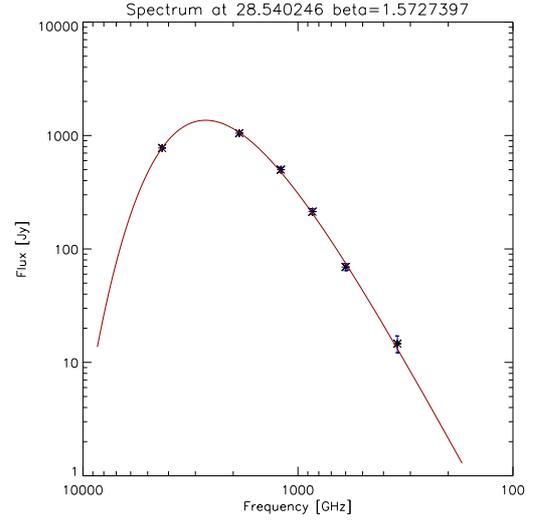}
 \caption{SED for \guno\ obtained using Herschel and ATLASGAL fluxes. }
  \label{sed1}
\end{figure}

We also estimated $T_{\rm dust}$ for \guno\   from the whole set of Herschel data and the LABOCA image by fitting a Planck function. Fluxes at 70, 160, 250, 350, and 500 $\mu$m were calculated with HIPE using rectangular sky aperture photometry and convolving all images down to the angular resolution  at 500 $\mu$m.  
For each aperture we measured the average surface brightness with an uncertainty derived from the standard deviation of the surface brightness within each aperture and the flux calibration uncertainty. Derived values are 776.0, 1047.2, 499.5, 213.3, and 69.3 Jy at 70, 160, 250, 350, and 500 $\mu$m, respectively. Fig.~\ref{sed1} shows the spectral energy distribution (SED), which also includes the flux density at 870 $\mu$m obtained by \citet{contre13} (14.63$\pm$2.47 Jy), and the best fitting (obtained using a greybody), which corresponds to T$_{\rm dust}$ = 28$\pm$1 K. Dust temperature for \gdos\ from a spectral energy distribution (SED) is difficult to determine. Due to the poor angular resolution of the Herschel images at 350 and 500 $\mu$m, fluxes estimated from these data includes part of S\,24, what can fake the dust temperature.

Dust masses can be estimated from the expression \citep{hild83}
\begin{equation}
{\rm M}_{\rm dust} = \frac{S_{\rm 870} \ d^{2}}{\kappa_{\rm 870} \ B_{870}(T_{\rm dust})}
\end{equation}
\noindent where $S_{\rm 870}$ is the flux density at 870 $\mu$m, $d$ = 3.7$\pm$0.5 kpc, $\kappa_{\rm 870}$ = 1.0 cm$^{2}$/gr is the dust opacity per unit mass \citep{oss94}, and $B_{\rm 870}(T_{\rm dust})$ is the Planck function for a temperature T$_{\rm dust}$. 

Two processes may contribute to the emission at this wavelength in addition to the thermal emission from cold dust: molecular emission from the CO(3-2) line and free-free emission from ionized gas. The continuum emission contribution at 345 GHz due to ionized gas was estimated from the radio continuum image at 1.4 GHz considering that the radio emission is thermal (see Sect. 5)  using the  expression $S_{\rm 345} = (345/1.4)^{-0.1} S_{\rm 1.4}$. Adopting $S_{\rm 1.4}$ = 1.2 Jy (see Sect. 5), this contribution amounts to  about 1\% of the emission at 870 $\mu$m. 
To estimate the contribution from the CO(3-2), we took into account the intensity in the $^{13}$CO(3-2) line. The  contribution of this mechanism is less than 1\%. Thus, the contribution by both mechanisms is within calibration uncertainties. 

Adopting $T_{\rm dust}$ = 28$\pm$1 K for \guno, and a  flux $S_{\rm 870}$ = 14.63$\pm$2.47 Jy derived by Contreras et al. (2013) for AGAL341.217-00.212, a dust mass $M_{dust1}$ = 12.9$\pm$6.3 \msun\ can be estimated for \guno. Assuming the same dust temperature for \gdos, and a flux $S_{\rm 870}$ = 25.55$\pm$4.11 Jy (for AGAL341.216-00.236), we estimate $M_{\rm dust2}$ = 22.4$\pm$11.0 \msun.  Finally, for the IRDC to the west of the bubble, we adopt temperatures in the range 10-20 K (see \citealt{peretto10}), which, along with a flux $S_{\rm 870}$ = 12.43$\pm$2.14 Jy for AGAL341.196-00.221, results in $M_{\rm dust-irdc}$ = 17-57 \msun. Uncertainties in these values are about  50\%. Adopting a gas-to-dust ratio of 100, gas masses are 1300 \msun\ for \guno, 2300 \msun\ for \gdos, and 1700-5700 \msun\ for the IRDC. 

%--------------------------------------------Fig. 7 
%-----------------------------------------------------

\section{The ionized gas}

%--------------------------------------------Fig. 8
\begin{figure}
  \centering  
\includegraphics[width=9cm]{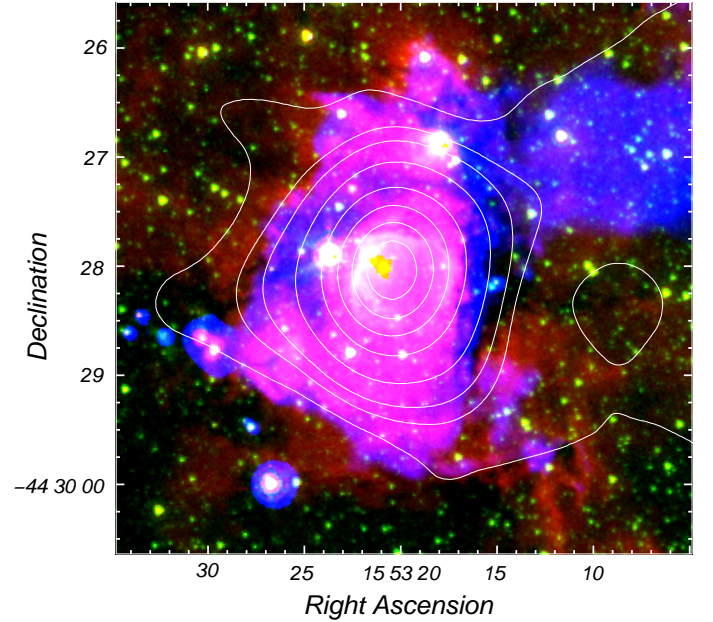}
 \caption{ Overlay of the radio continuum emission at 843 MHz (contours), the IRAC emission at 8 $\mu$m (in red), the MIPSGAL emission at 24 $\mu$m (in blue), and the IRAC emission at 3.6 $\mu$m (in green). Contours correspond to 10, 30, and 50 \mjyb, and from 100 to 500 \mjyb\ in steps of 100 \mjyb. }
  \label{cont-radio}
\end{figure}
%-----------------------------------------------------

Figure~\ref{cont-radio} displays an overlay of the radio continuum image at 843 MHz (in contours) and the emissions at 8 $\mu$m and 24 $\mu$m (in colorscale). 
The image at 843 MHz reveals radio emission coincident with the sources, and in particular with the S\,24 bubble. The emission at 1.4 GHz (not shown here) correlates with that at the lower frequency. We derived a flux density $S_{1.4}$ = 1.2 Jy.  Taking into account a flux density $S_{\rm 0.843}$ = 0.92 Jy, a spectral index $\alpha$ = 0.5 ($S_{\nu} \propto \nu^\alpha$) can be estimated from the emission at both frequencies. The $\alpha$-value indicates that at least at 843 MHz, the radio emission is optically thick, and that the source is thermal in nature. 

Figure~\ref{cont-radio} also shows that S\,24, as well as \guno, \gdos, and \gtres\ emit at 24 $\mu$m, indicating the presence of warm dust linked to the sources, and consequently, the existence of excitation sources inside them. Emission at this wavelength also coincides with the PDR, as revealed by the emission at 8 $\mu$m. In fact, the emission at 24 $\mu$m  from \hii\ regions has two components: emission from very small grains out of thermal equilibrium, detected inside the \hii\ region, and  thermal emission from the PDR from grains in thermal equilibrium \citep{anderson12}.

Lower limits for the electron density $n_e$ and the ionized mass $M_i$ can be derived from the image at 843 MHz. For an estimated flux density of 0.92 Jy, using the classical expressions by \citep{mezger67}, a typical  electron temperature of 10$^4$ K,  and bearing in mind a He abundance of 10\% by particle number and that He is single ionized, we derived $n_e \ >$ 200 \cmtres, and $M_i \ >$ 22 \msun.  We note that similar results can be obtained using the emission at 1.4 GHz. Clearly, the fact that the emission at 1.4 GHz is optically thick also can not be ruled out. Bearing in mind that the emission at 8 $\mu$m due to PAH in the photodissociation region marks the boundary of the ionized region, we measured a radius of 0.33 pc for the \hii\ region. 

Although these results can not be used to determine either the ionized mass or the electron density, they reveal the existence of an \hii\ region.

%--------------------------------------------Fig.9
\begin{figure*}
  \centering
  \includegraphics[width=18cm]{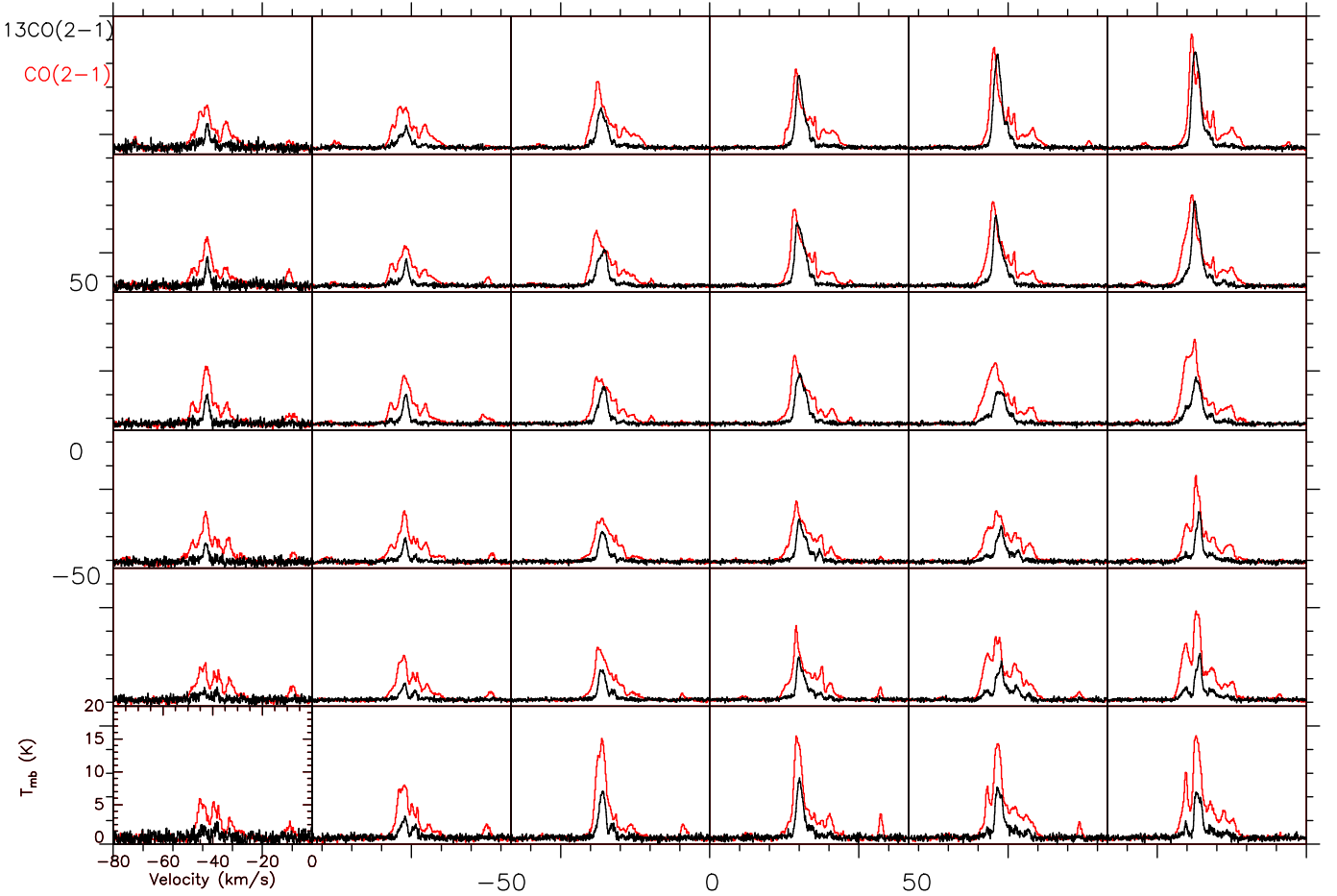}
 \caption{Decl. vs. R.A. map showing $^{12}$CO(2-1) (in red) and $^{13}$CO(2-1) (in black) profiles observed in a 5\arcmin $\times$5\arcmin\ region centered in the S\,24 bubble. Each profile shows T$_{\rm mb}$  in the interval from -1 K to 20 K vs. LSR velocity in the range from -80 \kms\ to 0 \kms. The (0,0) position corresponds to \radec\ = (16$^h$52$^m$20$^s$, --44\degr 27\arcmin 30\arcsec). Note that R.A. inceases towards the right.  }
  \label{12spectra}
\end{figure*}
%-----------------------------------------------------

%--------------------------------------------Fig. 10
\begin{figure}
  \centering
  \includegraphics[width=9cm]{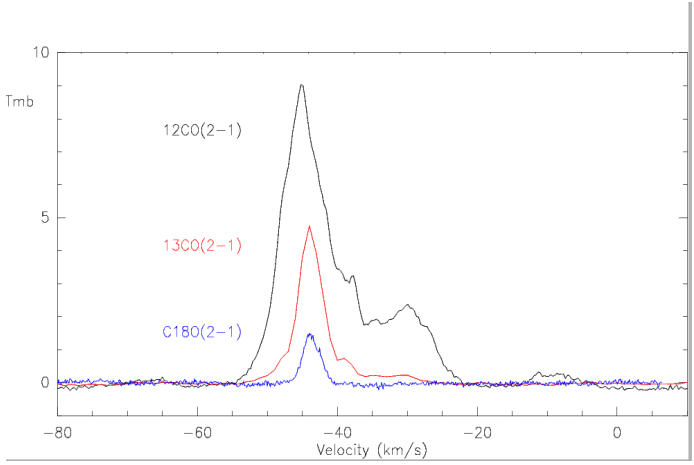}
 \caption{CO(2-1) (in black),  $^{13}$CO(2-1) (in red), and  C$^{18}$O(2-1) (in blue) spectra  obtained by averaging the observed spectra in the region of S\,24.
Intensity is expressed as main-beam brightness-temperature. 
}
  \label{co-spectra}
\end{figure}
%------------------------------------------------------

\section{The distribution of the molecular gas}

Figure~\ref{12spectra} displays the CO(2-1) and $^{13}$CO(2-1) spectra obtained for the region under study. The spatial separation between these profiles is 30\arcsec. Relative coordinates are expressed in arcseconds, referred to \radec\ = (16$^h$52$^m$20$^s$, --44\degr 27\arcmin 30\arcsec). The individual spectra show that  the molecular emission appears concentrated  between $\sim$--55 \kms\ and --20 \kms, where a number of velocity components are present, and an isolated component near --10 \kms\ (which is not going to be taken into account). The  CO(2-1)/$^{13}$CO(2-1) line ratios are close to 1-1.5 in the central and northeastern parts of the figure, suggesting that the gas is optically thick at $^{13}$CO(2-1) where S\,24, \guno, and \gdos\ are located. 

The CO spectra of Fig.~\ref{co-spectra} illustrate the general behaviour of the molecular gas towards the entire region. There, we show the  CO(2-1), $^{13}$CO(2-1), and  C$^{18}$O(2-1) spectra averaged within the  observed region. At least three velocity components can be identified in the line of sight within the velocity interval from --55 \kms\ to -20 \kms.  

The bulk of the molecular emission is detected in the three molecular lines and peaks at $\simeq$--45 \kms, while fainter gas components are present at  $\simeq$--38 \kms\ and --30 \kms. The faintest component is present near --10 \kms\ in the $^{12}$CO(2-1) line only. The velocity components peaking at --38 \kms\ and --30 \kms\ can be barely identified in the $^{13}$CO(2-1) spectrum and are absent in the C$^{18}$O(2-1) profile. Gas components having the same velocity can be also identified in the averaged J = 3-2 profile (not shown here).

Circular galactic rotation models (e.g. \citealt{b&b93}) predict near and far kinematical distances in the range 3.2-3.9 kpc and 12-13 kpc, respectively, for the gas component at --45 \kms, while near and far kinematical distances in the range 2.4-3.5 and 12-14 kpc, respectively, were obtained for gas in the  components at --38 \kms\ and --30 \kms. Particularly for the component at --45 \kms, near kinematical distances are compatible with previously adopted values for the region of S\,24.

%--------------------------------------------Fig.11
\begin{figure*}
  \centering
  \includegraphics[width=15cm]{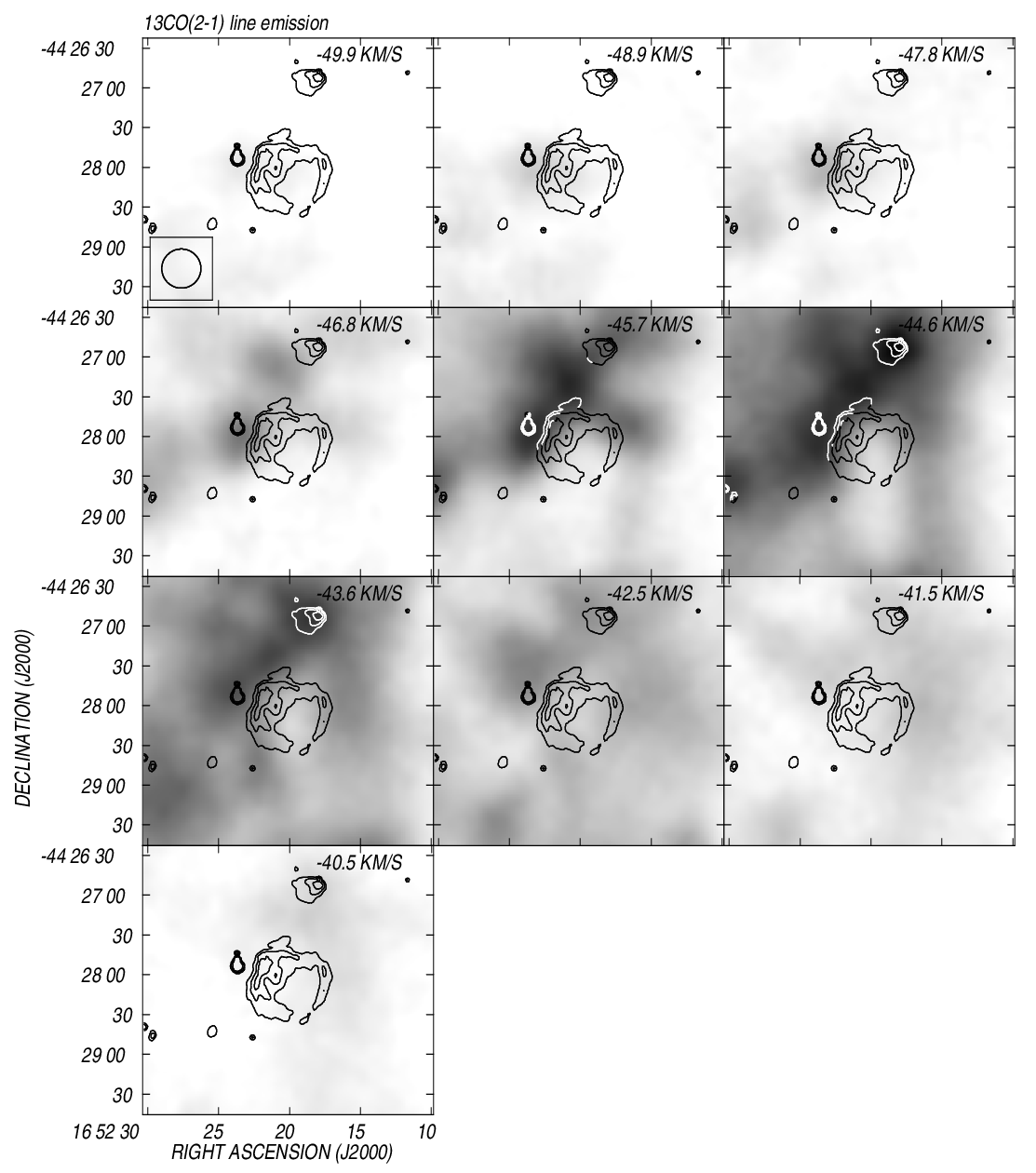}
 \caption{ Channel maps of the $^{13}$CO(2-1) line (in grayscale) for the velocity interval from --49.9 \kms\ to --40.5 \kms\ in steps of 1.05 \kms\ overlaid onto the 8 $\mu$m emission (in contours). Grayscale goes from 0.5 K to 17.0 K.  }
  \label{maps}
\end{figure*}
%-----------------------------------------------------

To illustrate the spatial distribution of the molecular gas we show in Fig.~\ref{maps} the emission of the $^{13}$CO(2-1) in the interval --49.9 \kms\ to --40.5 \kms, in steps of 1.05 \kms\ (in grayscale) superimpossed to the emission at 8 $\mu$m (in contours) for comparison. Most of the molecular emission has velocities in the interval from --46.8 \kms\ to --41.5 \kms. 

%--------------------------------------------Fig.12
\begin{figure*}
  \centering
  \includegraphics[width=9cm]{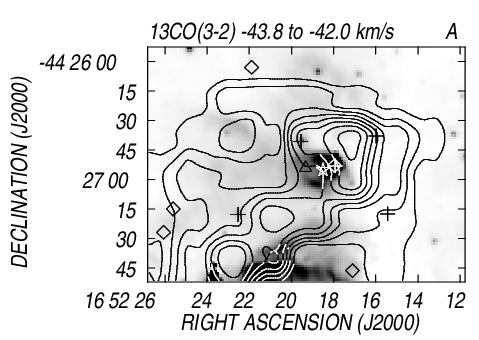}
  \includegraphics[width=9cm]{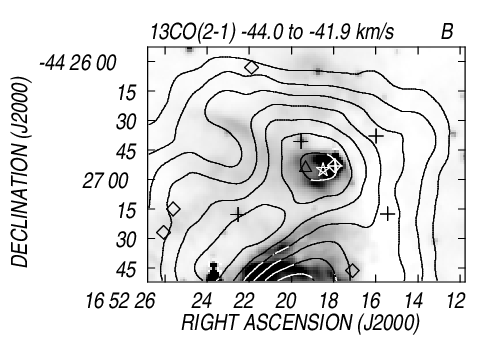}
  \includegraphics[width=9cm]{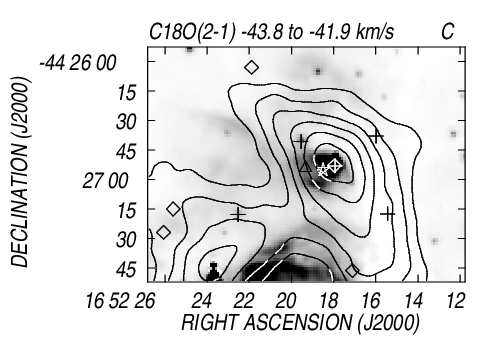}
  \includegraphics[width=9cm]{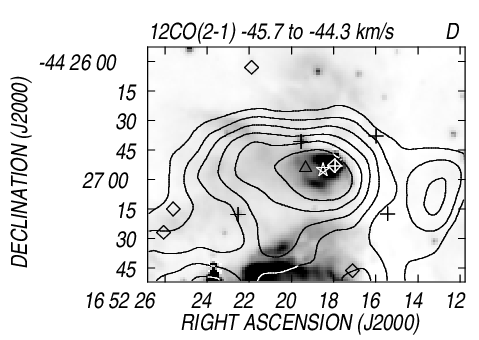}
 \caption{{\it Panels A-D:} Overlay of the CO emission distribution in contours and the emission at 8 $\mu$m in grayscale towards \guno. {\it Panel A:} $^{13}$CO(3-2) line emission in the interval from --43.8 \kms\ to --42.0 \kms. Contours are from 5 K to 12 K ($T_{\rm mb}$) in steps of 1 K. the different symbols mark the position of YSO candidates: {\it stars:} MSX sources, {\it triangles:} 2MASS sources, {\it crosses:} Spitzer sources, {\it diamonds:} WISE sources. {\it Panel B:} $^{13}$CO(2-1) line emission in the interval from --44.0 \kms\ to --41.9 \kms. Contours are from 5 K to 11 K ($T_{\rm mb}$) in steps of 1 K. {\it Panel C:} C$^{18}$O(2-1) line emission in the interval from --43.8 \kms\ to --41.9 \kms. Contours are from 1.5 K to 4 K ($T_{\rm mb}$) in steps of 0.5 K. {\it Panel D:} $^{12}$CO(2-1) line emission in the interval from --45.7 \kms\ to --44.3 \kms. Contours are from 9 K to 14 K ($T_{\rm mb}$) in steps of 1 K. }
  \label{g34121}
\end{figure*}
%-----------------------------------------------------

 %---------------------------------figure 13
\begin{figure*}
\centering
\includegraphics[width=8cm]{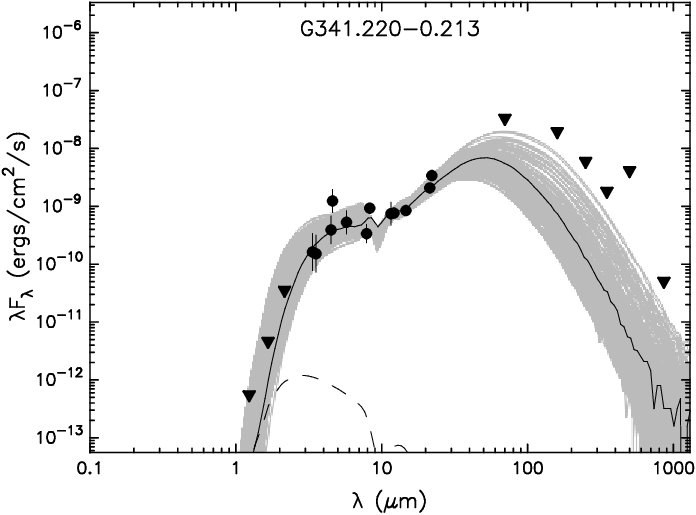}
\includegraphics[width=8cm]{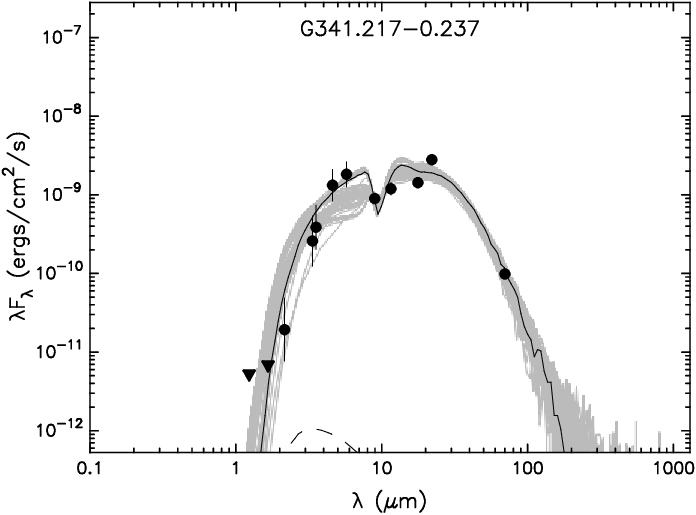}
\caption{{\it Left panel:} SED of YSOs  from Table 1 in \guno. Input data are indicated by filled circles. The triangles indicate upper limits. The black line corresponds to the best fitting. The fittings which obey Eq.~1 appear indicated by gray lines. The dashed line shows the emission of the stellar photosphere including foreground interstellar extinction. {\it Right panel:} SED of  source \#43 from Table 1 in \gdos.  }
\label{sed-yso}
\end{figure*}
%---------------------------
%--------------------------------------------Fig.14
\begin{figure*}
  \centering
 \includegraphics[width=13cm]{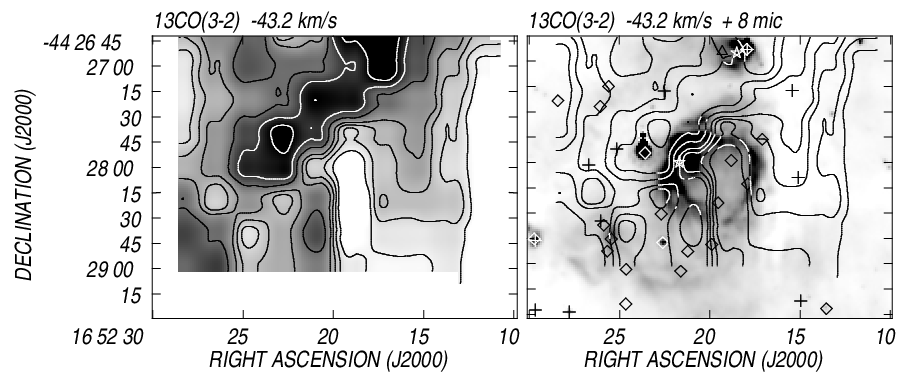}
  \includegraphics[width=13cm]{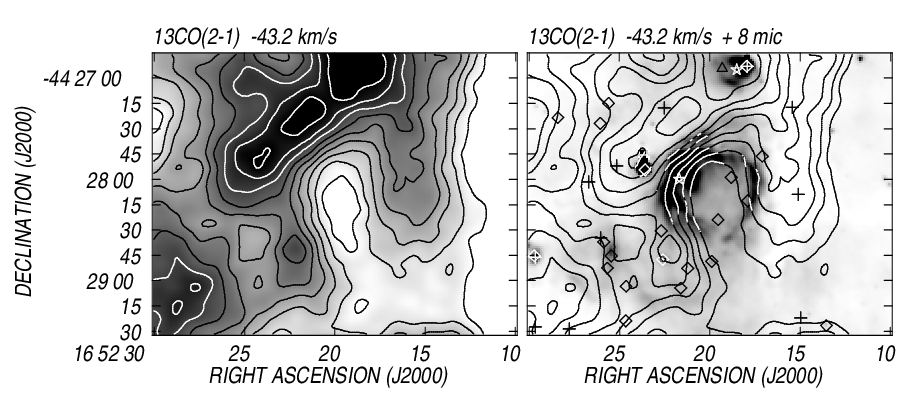}
  \includegraphics[width=13cm]{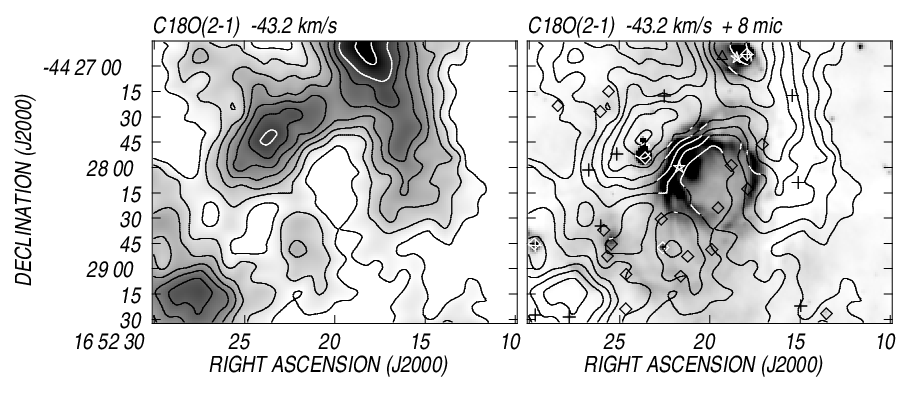}
 \caption{ {\it Upper panels:} {\it Left:} $^{13}$CO(3-2) emission distribution at --43.2 \kms\ showing the molecular envelope around the S\,24 bubble. Contours are 3.0 K to 6.0 K ($T_{\rm mb}$), in steps of 1.0 K; 8.0 K, 10.0 K, and 12.0 K.  {\it Right:} Overlay of the same contours of the left panel and the IRAC emission at 8 $\mu$m. The symbols have the same meaning as in Fig. 12.  
{\it Middle  panels:} {\it Left:} $^{13}$CO(2-1) emission distribution at --43.2 \kms. Contours are from 4.0 K to 12.0 K ($T_{\rm mb}$), in steps of 1.0 K.  {\it Right:} Overlay of the same contours of the left panel and the IRAC emission at 8 $\mu$m.   
{\it Bottom  panels:} {\it Left:} C$^{13}$O(2-1) emission distribution at --43.2 \kms. Contours are from 1.0 K to 4.0 K ($T_{\rm mb}$), in steps of 0.5 K, and 5 K.  {\it Right:} Overlay of the same contours of the left panel and the IRAC emission at 8 $\mu$m.    
 }
  \label{cascara}
\end{figure*}
%-----------------------------------------------------

As regards the S\,24 bubble, CO emission encircles the eastern and northern sections at --46.8 \kms\ and --45.7 \kms. In the images at --44.6 \kms, --43.6 \kms, and --42.5 \kms\ this emission extends to the western section of the bubble, where the IRDC SDS \hbox{341.194-0.221} is present. S\,24  appears almost completely encircled by molecular gas in the images at --43.6 \kms\ and --42.5 \kms, while at --41.5 \kms\ and --40.5 \kms\ the CO emission is present to the west of the bubble, coincident with the IRDC. 

A close inspection of the emission in the interval from --48.9 \kms\ to --42.5 \kms\ reveals bright molecular emission linked to \gdos\ and the eastern border of the IR bubble.       

The CO emission associated with the different IR sources described above will be analyzed in more detail in the following sections. Material linked to the IRDC SDC341.232-0.268 and present in the interval from --49.5 \kms\ to --42.1 \kms\ is being analyzed in detail by Vasquez et al. (in preparation). 

\section{Putting all together}

In this section we investigate in more detail the gas distribution in S\,24 and the different features in its environment and its relation to the warm and cold dust and the YSO candidates. 

%--------------------------------------------Fig.15
\begin{figure}
  \centering
  \includegraphics[width=7cm]{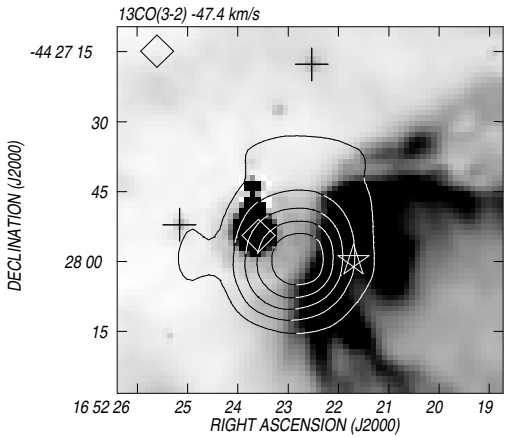}
 \caption{Overlay of the  $^{13}$CO(3-2) emission distribution at --47.4 \kms\ (in contours) and the emission at 8 $\mu$m (in grayscale) showing the molecular clump at the interface between \gdos\ and the S\,24 bubble. Contours are from 3 K to 6  K ($T_{\rm mb}$), in steps of 1 K. The symbols are the same as in Fig. 12.
 }
  \label{clump}
\end{figure}
%-----------------------------------------------------

\subsection{\guno}

Panel A of Fig.~\ref{g34121} displays the $^{13}$CO(3-2) emission having velocities in the range --43.8 \kms\ and --42.0 km s$^{-1}$ superimposed onto the emission at 8 $\mu$m for comparison. The emission distribution at these velocities reveals a molecular arc-like structure closely encircling the bright  northern and western borders of \guno. At 8 $\mu$m, \guno, which  resembles a horseshoe, is open opposite to the molecular emission. The IR source is clearly interacting with the surrounding molecular gas  and is evolving in a medium with a density gradient. The coincidence of \guno\ with emission at 24 $\mu$m (see Sect.~5) suggests the presence of excitation sources inside. Bearing in mind the morphology of the source at 8 $\mu$m and in molecular lines,  a champagne flow \citep{garay99} can not be discarded.

A number of YSO candidates appears inside the horseshoe: one MSX source classified as MYSO (\# 2 in Table 1); one Spitzer and one WISE sources coincident in position, both  classified as Class I (\# 15 and \# 34). We inspected the spectral energy distribution (SED) of  \# 2, \# 15, and \# 34, assumed to be only one source, using the on-line\footnote{http://caravan.astro.wisc.edu/protostars/} tool developed by \citet{robitaille07}, which can help to discriminate between evolved stars and reliable candidate YSOs. This tool fits radiation transfer models to observational data according to a $\chi^2$ minimization algorithm. Models that accomplished the following condition were selected:
\begin{equation}
\chi ^2 \ - \ \chi ^2_{\rm min} \ < \ 3n
\end{equation}
\noindent where $\chi^2_{\rm min}$ is the minimum value of the $\chi^2$ among all models, and $n$ is the number of input data fluxes.

To perform the fitting we used the photometric data listed in Table 1, along with fluxes derived from Herschel and ATLASGAL images, visual absorption values in the range 30-40 mag, and distances in the range from 3.2 to 4.2 kpc. The result is shown in the left panel of Fig.~\ref{sed-yso}. Fluxes obtained from Herschel and ATLASGAL were considered upper limits. The best fitting suggests a massive central source of  about 10 \msun, a disk mass of 5$\times$10$^{-4}$ \msun, an envelope mass of 21 \msun, a total luminosity of 5300 \lsun, and an age of 1$\times$10$^6$ yr. With this characteristics, the evolutionary stage of the central source according to \citet{robitaille07} would correspond to Stage II  (objects with optically thick disks). These results points to a massive central object. 

We propose that this source has started modifying its environs by dissociating and then ionizing the nearby material, originating a PDR detected by its emission at 8 $\mu$m at the interface between the ionized and molecular material. It may have contributed to the heating of the dust as revealed by the emission at 24 $\mu$m. Also, the detection of maser emission towards \guno\ (see Sect. 1) indicates that star formation is active.

A low emission region is present at  \radec\ = (16$^h$52$^m$22.5$^s$, --44\degr 26\arcmin 40\arcsec), probably linked to a weak IR arc seen slightly to the north of that position.

Panels B and C of Fig.~\ref{g34121} show the emission distribution in the $^{13}$CO(2-1) and C$^{18}$O(2-1) lines for the same velocity interval as panel A, while in panel D, the emission corresponds to the  CO(2-1) line in a slightly different velocity interval. The emission distribution reveals a molecular clump coincident with the \guno. No signs of the arc-like structure are present in the (2-1) transitions, probably due to the lower angular resolution of these lines (30\arcsec) in comparison with the one of the $^{13}$CO(3-2) (20\arcsec).   

The molecular clump detected in the interval [--44.0,--42.0] \kms\ coincides with the emission of cold dust as imaged in the far-IR by LABOCA (AGAL~341.217-00.212) and Herschel (see Fig.~\ref{herschel}), indicating that gas and dust are well mixed in the region. 

The analysis of the data cubes allows us to identify molecular gas associated with \guno\ in the velocity interval from --46.5 \kms\ to --40.2 \kms.

\subsection{The S\,24 \hii\ region}

The upper left panel of Fig.~\ref{cascara} shows the $^{13}$CO(3-2) emission distribution at  --43.2 \kms\ in grayscale and contours, while the upper right panel displays an overlay of the same CO contours and the emission at 8 $\mu$m. The middle and bottom panels show the emission at the same velocity for the $^{13}$CO(2-1) and C$^{18}$O(2-1) lines, respectively.

The emission at this velocity reveals an U-shaped, almost complete molecular shell encircling the S24 bubble. The strongest emission regions of this molecular shell encircles the brightest sections  of S\,24 as seen at 8 $\mu$m. The shell appears open towards  \radec\ = (16$^h$52$^m$19$^s$, --44\degr 28\arcmin 30\arcsec). The emission in the western part of the shell coincides with the IRDC SDC341.194-0.221, which borders this section of the bubble.  The molecular shell is present in the interval from --46 \kms\ to --41.5 \kms.

As shown in Fig.\ref{maps}, in the velocity interval from \hbox{--46.5} \kms\ to --43.5 \kms\ the molecular gas borders the eastern, northern and western sections of S\,24. Molecular gas at these velocities coincides with cold dust detected in the far IR at $\lambda >$ 160 $\mu$m, as shown in Figs. 2 and 4. Molecular shells have been found around many IR dust bubbles, in particular those identified as \hii\ regions (e.g. \citealt{petriella10,pomares09,deharveng10}). The presence of CO in the \hii\ region interior suggests that the shell corresponds to a three-dimensional structure. The appearance of the shell indicates that it is unhomogeneous, with the molecular gas distributed unisotropically (e.g. \citealt{anderson15}). 

The molecular shell might be  expanding at 2-3 \kms\ due to the difference in pressure between the ionized gas inside the bubble and the neutral gas outside, although no signs of expansion are detected in position-velocity diagrams. We note, however, that low expansion velocities as those expected in this case are not easy to detect. In the case of expansion, the emission detected towards the interior of the region would correspond to the approaching and receding sections of the shell. 

Three YSO candidates are projected onto the S\,24 bubble: one MYSO (\# 1 in Table 1), and two WISE sources identified as Class II (\# 35 and \# 36). 

Also, a number of YSOs coincide in position with the molecular shell, suggesting that star formation is active in the environs of the bubble. 

Two YSOs appear projected onto the IRDC SDC341.194-0.221. They do not coincide with the spot of maser emission described in Sect.\,1 projected onto this IRDC.

In addition to these YSO candidates, three stars could be identified in the 2MASS catalog in the region under study. These stars are indicated with green asterisks in the [H-K$_s$, J-H] and [H-K$_{\rm s}$, K$_{\rm s}$ ] diagrams of Fig.~\ref{cc-diagrams}. In particular, 2MASS J17502070-4428012 is projected inside the bubble at \radec\ = (16$^h$52$^m$20.7$^s$, --44\degr 28\arcmin 1\farcs 2). It locus in the CC- and CM diagrams (H-K$_s$) = 1.19 mag, (J-H) = 2.18 mag,  and K$_{\rm s}$ = 10.94 mag) is indicative of a massive star with an optical absorption $A_{\rm V} \simeq$ 30 mag. 
Although more studies are neccessary, this star might be responsible for dissociating and ionizing the gas, creating the observed PDR at 8 $\mu$m.

%-------------------------------------Figure 16
\begin{figure*}
\centering
\includegraphics[width=370pt]{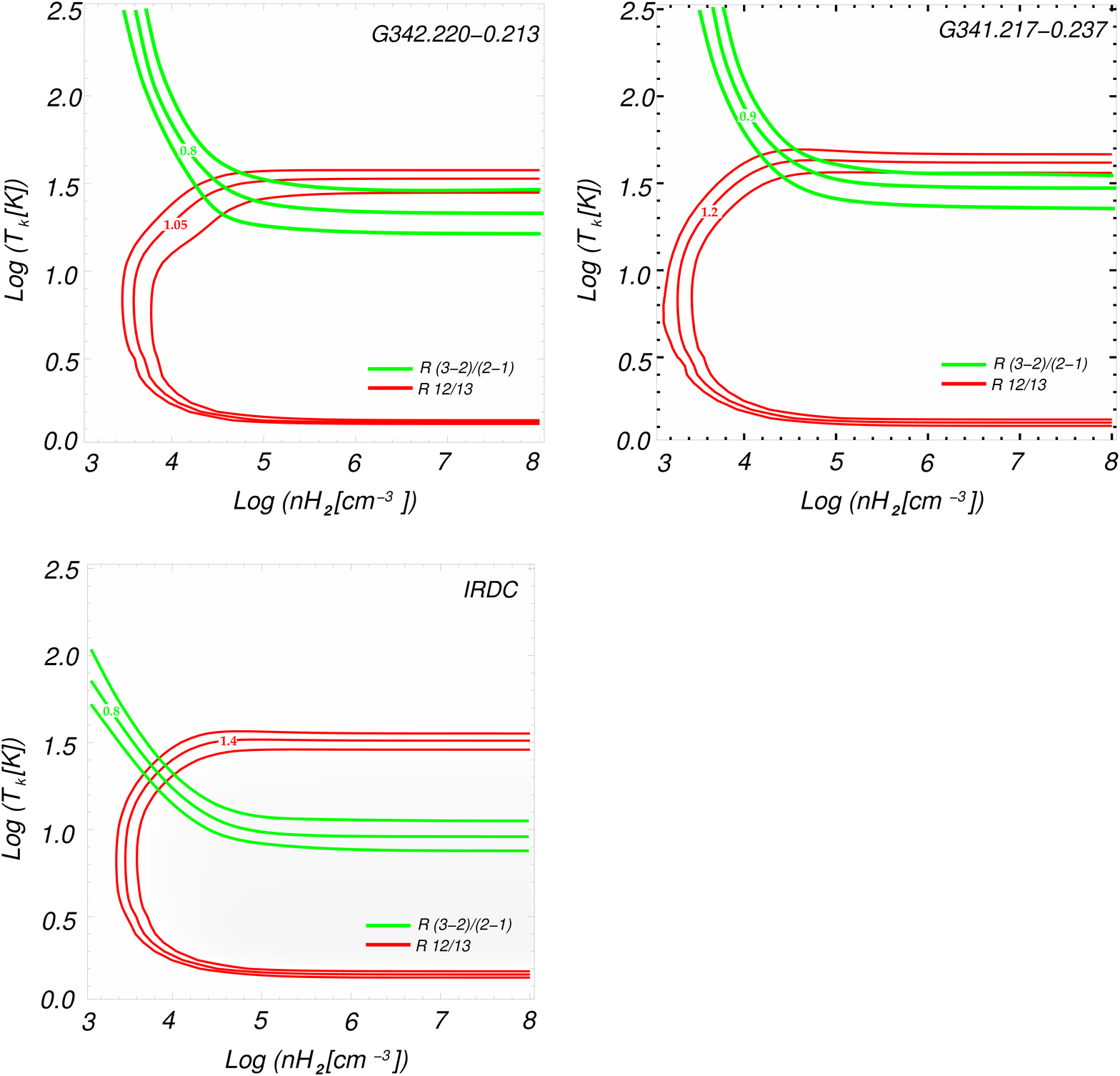}
\caption{ LVG plots  for G341.220-0.213, G341.217-0.237, and the IRDC SDC341.194-0.221. Green lines mark the solutions for the ratios $^{13}$CO(3-2)/$^{13}$CO(2-1),  while red lines, for the ratios $^{13}$CO(2-1)/CO(2-1). Central lines correspond to the observed ratios. Outer lines indicate $\sim$10$\%$ calibrations uncertainties.  }
\label{lvg}
\end{figure*} 
%---------------------------------------

\subsection{The interface between S\,24 and \gdos}

As described in Sect. 6, a bright molecular clump appears projected onto the interface between \gdos\ and S\,24 at \radec\ = (16$^h$52$^m$23$^s$, --44\degr 28\arcmin). Figure~\ref{clump} displays an enlargement of the region of the clump showing the emission in the {\bf $^{13}$CO(3-2)} line at --47.4 \kms\ (in contours) and the 8 $\mu$m image in grayscale. The peak CO emission is coincident  with the interface between the two sources.  
The anticorrelation between the CO and IR emissions  strongly suggests that the molecular clump has modeled the eastern border of the PDR in the \hii\ region. 

These facts suggest that S\,24 and \gdos\ are interacting with the molecular clump. The molecular clump coincides with emission at 870 $\mu$m and  in the Herschel bands at 250 and 350 $\mu$m. 

The YSO candidate \# 43, classified as Class I, coincides with \gdos. The SED corresponding to this source is shown in the right panel of Fig.~\ref{sed-yso}. To perform the fitting we used the photometric data listed in Table 1, and the flux derived from the image at 70 $\mu$m. Herschel fluxes at larger wavelengths include a significant part of the S\,24 bubble and were not used for the SED. Visual absorption is in the range 30-50 mag, and distances from 3.2 to 4.2 kpc. The best fitting corresponds to a central source of  about 15 \msun, a disk mass of 8$\times$10$^{-3}$ \msun, an envelope mass of 1.1$\times$10$^{-5}$ \msun, a total luminosity of 22000 \lsun, and an age of 1.3$\times$10$^6$ yr. According to this fitting, the object is probably in Stage III (objects with optically thin disks). However, as pointed out by \citet{deharveng12}, \citet{offner12}, and \citet{robitaille08}, these results should be taken with caution.
The presence of emission at 24 $\mu$m coincident with \gdos\ suggests that  the ionizing photons from the massive source are heating the dust and probably dissociating and ionizating the molecular gas.

\section{The parameters of the molecular gas}

The molecular mass linked to \guno, the clump at the interface between \gdos\ and S\,24,  and the S\,24 \hii\ region, and the molecular mass in the whole region were  estimated from the emission in the J=2$\rightarrow$1 line. The derived parameters are listed in Table~\ref{masses}.

Assuming Local Thermodinamic Equilibrium (LTE) conditions and that the emission in the CO(2-1) line is optically thick, we derived the  excitation temperature T$_{exc}$ from the emission in the CO(2-1) line {\bf using 
\begin{equation}
T_{\rm p-12CO} = T_{\rm 12CO}^*\left[\left(e^{\frac{T_{12}^*} {T_{\rm exc}}}-1\right)^{-1}
-\left(e^{\frac{T_{12}^*} {T_{\rm bg}}}-1\right)^{-1}\right]
\label{texc12co}
\end{equation}
where $T_{\rm 12}^*$ = $h \nu_{\rm 12} / k$, $\nu_{\rm 12}$ the frequency of the $^{12}$CO(2-1) line,  and $T_{\rm bg} = 2.7$ K. 
To obtain the peak main beam temperature in the CO(2-1) line $(T_{\rm p-12CO})$ we took into account the averaged spectrum within regions of about 30\arcsec\ in diameter for the cases of \guno\ and the interface between \gdos\ and S\,24, while for the S\,24 region and the whole region the spectra were averaged in a region of 2\arcmin\ and 3\arcmin\ in diameter, respectively.  }

The optical depth $\tau_{13}$ was obtained from the $^{13}$CO(2-1) line  by assuming that the excitation temperature is the same for CO(2-1) and $^{13}$CO(2-1) emission lines using the expression 
\begin{equation}
\tau_{13} = -{\rm ln}\left[1-\frac{ T_{\rm p-mb}(^{13}CO)}{T_{13}^*}
\left[\left(e^{\frac{T_{13}^*} {T_{\rm exc}}}-1
\right)^{-1}-\left(e^{\frac{T_{13}^*} {T_{\rm bg}}}-1\right)^{-1}\right]^{-1}\right]
\label{tau13co}
\end{equation}
where $T_{\rm 13}^*$ = $h \nu_{\rm 13} / k$, $\nu_{\rm 13}$ the frequency of the $^{13}$CO(2-1) line. 
Assuming LTE,  the $^{13}$CO column density, $N_{tot}(^{13}CO)$, can be estimated from the $^{13}$CO(2-1) line data following
\begin{equation}
N_{\rm tot}(^{13}CO) = 3.2\times 10^{14}\ \left[\frac{e^{T_{13}^*/T_{\rm exc}}}{1 -e^{-T_{13}^*/T_{\rm exc}} }\right] \ T_{\rm exc} \int \tau^{13}\ \ dv  \textrm{ (cm$^{-2}$)}
\label{n13co}
\end{equation}
The integral of Eq. \ref{n13co} can be aproximated by
\begin{equation}
T_{exc} \int{\tau^{13} dv \approx\ \frac{\tau^{13}}{1-e^{(-\tau^{13})}}
\int{T_{\rm mb}}}\ \ d{\rm v}.
\label{integral}
\end{equation}
This aproximation helps to eliminate to some extend optical depth effects and is good within 15\% for $\tau <$  2 (\citealt{rw04}). Bearing in mind the $\tau_{\rm 13}$-values listed in Table 2, it is appropriate for our region.  Then, the molecular mass was calculated using
\begin{equation}\label{eq:masa}
{\rm M(H_2)}\ =\ (m_{sun})^{-1}\ \mu\ m_H\  A \ {\it N}(\rm H_2)\ {\it d}^2 \quad \quad \quad \textrm{(M$_{\odot}$)}
\end{equation}
where m$_{\rm sun}$ is the solar mass ($\sim$2$\times$10$^{33}$ g), $\mu$ is the mean molecular weight, which is assumed to be equal to 2.76 after allowance of a relative helium abundance of 25\% by mass \citep{Y99}, m$_{\rm H}$ is the hydrogen atom mass ($\sim$1.67$\times$10$^{-24}$ g),  $A$ is the solid angle of the CO emission, $d$ is the adopted distance expressed in cm, and $N$(H$_{\rm 2}$) is the H$_{\rm 2}$ column density, obtained using an abundance \hbox{$N(\rm H_2)$ / $N(^{13} {\rm CO})$} = \hbox{5$\times$10$^{5}$} \citep{dickman78}. Uncertainties in molecular masses and ambient densities are about 50\%\ and 70\%, respectively, and originates mainly in distance uncertainties and optically thick emission in the $^{13}$CO(2-1) line.

%-------------------------------------------Table --------------
\begin{table*}
\centering
\caption[]{Molecular gas parameters }
\begin{tabular}{lcccccccc}
\hline
\hline
%\multicolumn{3}{l}{\bf Molecular gas parameters:}\\
  & I($^{13}$CO) & T$_{\rm exc}$ &$\tau_{\rm 13CO}$ & R & $N(^{13}CO)$ & $N({\rm H_{\rm 2}})$ & $M_{\rm mol}$ &  $n_{\rm H_{2}}$\\
&  [K \kms]&    & [K] & [\arcsec = pc] & [10$^{16}$ cm$^{-2}$] & [10$^{22}$ cm$^{-2}$]  & [\msun] &  [10$^4$ cm$^{-3}$]\\
\hline
G341.220-0.213  & 52.3  & 19.1 & 1.95 &  27 = 0.44 & 14.6  & 7.3 &  1550 & 6.6 \\
Interface G341.217-0.237/S24  & 18.9  & 21.0 & 1.65 &  10 = 0.16 &  5.2  & 2.6 &  470  & 41.6 \\
S\,24-shell     & 34.4  & 16.8 & 1.12 &  48 = 0.77 &  7.3  & 3.6 &  3100 & 2.5 \\
Whole region    & 32.4  & 16.1 & 1.14 & 114 = 1.8  &  7.0  & 3.5 & 10300 & 0.59 \\
\hline
\hline
\end{tabular}
\label{masses}
\end{table*}
%--------------------------------------------------------

To explore more robustly the issues of temperature and density on G341.220-0.213 and  G341.217-0.237 we performed a large velocity gradient (LVG) analysis \citealt{scoville73,goldreich74}) for radiative transfer of molecular emission lines. We use the {\it lvg} task implemented as part of the MIRIAD\footnote{http://www.cfa.harvard.edu/sma/miriad/packages/} package of SMA. For a given  column density (normalized by the line width), this program estimates the line radiation temperature  of a molecular transition as a function of the kinetic temperature ($T_{\rm k}$) and the H$_2$ volume density ($n_{H2}$). We generated  50 $\times$ 50 model grids for the  $^{13}$CO(3-2),  $^{13}$CO(2-1), and CO(2-1) lines  with T$_{\rm k}$ in the range 1-300 K and $n_{H2}$ with the range 1$\times$10$^3$ to 1$\times$10$^8$ cm$^{-3}$.   In all cases we convolved the $^{13}$CO(3-2) data with a Gaussian  of FWHM 30$''$ in order to smooth the cubes down to the {\bf angular} resolution  the  $^{13}$CO(2-1) {\bf data}. We then derived peak main beam temperatures and line widths for the studied components by fitting Gaussian functions to the line profiles. Two ratios, R$_{\rm (3-2)/(2-1)}$ = $T_{\rm mb}$$^{13}$CO(3-2)/CO(2-1) and R$_{\rm 12/13}$=$T_{\rm mb}$$^{13}$CO(2-1)/CO(2-1), are then calculated. The input column densities of CO and $^{13}$CO were estimated from  the dust-derived column density measured from the 870 $\mu$m emission in each region, and assuming relative abundances N(CO)/N(H$_2$) = 1$\times$10$^{-4}$ and N($^{13}$CO)/N(H$_2$) = 2$\times$10$^{-6}$ \citep{dickman78}. The same procedure was applied to the IRDC SDC341.194-0.221, for which we have not derived molecular mass and ambient density since its molecular emission is included in the S\,24 molecular shell.

Fig.\ref{lvg} shows the solutions for G341.220-0.213, G341.217-0.237, and the IRDC. They indicate that kinetic temperatures in the IRDC are about 20 K, not different than derived in other IRDCs (\citealt{egan98}), and not much higher than typical temperatures  of molecular gas without an extra heat source other than interstellar radiation field. The volume density of the IRDC is about 10$^4$ cm$^{-3}$. For the case of G341.220-0.213, G341.217-0.237, their temperatures are a bit higher ($\sim$ 30 - 40 K), not surprising given that they are active star-forming regions (see Sect. 7).  Regarding their volume densities, they are less well constrained but values higher than 4$\times$10$^4$ cm$^{-3}$ are inferred from the plots.

An inspection of Table 2 shows that molecular gas densities are in the range 10$^4$-10$^5$ for particular regions including the shell around the S\,24 \hii\ region. Masses derived from molecular data are compatible, within errors, with those obtained from the emission at 870 $\mu$m, assuming a gas-to-dust ratio equal to 100. Additionally, molecular ambient densities are compatible with values derived from LVG analysis. The large difference for \gdos\ arises in the fact that the mass estimate from molecular emission corresponds only to the emission at the interface between \gdos\ and the S\,24 \hii\ region.  The ambient density  obtained by distributing the total molecular mass in the whole observed region is high (5900 \cmtres), indicating that \guno, \gdos\, and S\,24 are evolving in a high density ambient medium.

The high opacity estimated for \guno\ is not surprising bearing in mind the high column density of this source.

\section{Triggered star formation scennario in S\,24?}

We have found evidences for ongoing star formation in the dense molecular envelope encircling the S\,24 \hii\ region. Have these YSOs been triggered through the {\it collect and collapse (CC)} mechanism? To answer this question we can apply the analytical model by \citet{whitworth94}. For the case of \hii\ regions, the model predicts the age of the \hii\ region at which the fragmentation occurs (the fragmentation time scale), $t_{\rm frag}$, the size of the \hii\ region at that moment, $R_{\rm frag}$, the mass of the fragments, $M_{\rm frag}$, and their separation along the compressed layer, $r_{\rm frag}$. The parameters required to derive these quantities are the UV photon flux of the exciting star, $N_{\rm Ly}$, the ambient density of the surrounding medium into which the \hii\ region is evolving, n$_{0}$, and the isothermal sound speed in the shocked gas, $a_{\rm s}$.

Since the exciting star is unknown, we took into account a large range of spectral types, i.e. from O5V to O9V stars, with UV fluxes in the range $N_{\rm Ly}$ = (18-0.8)$\times$10$^{48}$ s$^{-1}$ \citep{martins02}. Using the mean  $H_{\rm 2}$ ambient density $n_{H2}$ = 5900 \cmtres\ (see Table 2), and $a_{\rm s}$ = 0.2-0.6 \kms,  we obtained $t_{\rm frag}$ = (6.6-8.8)$\times$10$^5$ yr, $R_{\rm frag}$ = 1.8-2.3 pc, $M_{\rm frag}$ = 22-29 \msun, and, $r_{\rm frag}$ = 0.5-0.4 pc. 

The dynamical age of the \hii\ region can be estimated using the equation \citep{dw97}
\begin{equation}
t_{\rm dyn} \ = \frac{4 R_{\rm S}}{7 c_{\rm s}} \left[ \left(\frac{R}{R_{\rm S}}\right)^{7/4} \ -1\right]
\end{equation}
\noindent where $R_{\rm S}$ is the original Str\"omgren radius, equal to 0.09-0.25 pc for the adopted spectral types, and $c_{\rm s}$ is the sound velocity in the ionized gas. Derived dynamical ages span the range (9-43)$\times$10$^3$ yr. We find that the dynamical age is smaller than the fragmentation time scale  $t_{\rm frag}$ for the adopted ambient density. The comparison of these two quantities does not support a {\it collect and collapse} process, which is a relatively slow process, for the triggering of star formation in the envelope. An RDI scenario could be investigated, however evidences of this process (such the presence of pillars) appear to be absent. Very probably, the compact \hii\ region is too young for triggering to have begun.

\section{Conclusions}

Based on the molecular emission in the CO(2-1), $^{13}$CO(2-1),  C$^{18}$O(2-1), and $^{13}$CO(3-2) lines, obtained with the APEX telescope, and images in the near-, mid-, and far-IR from IRAC-Glimpse, Herschel, and ATLASGAL we performed a multi-wavelength analysis of the infrared dust bubble S\,24, and two extended IR sources (G341.220-0.213 and G341.217-0.237) located in its environs. The region coincides with the IRAS point source 16487-4423, classified as UCHII. We also investigated the presence of YSO candidates in the region using the MSX, 2MASS, Spitzer, and WISE catalogs.

The molecular emission  distribution shows that gas linked to the S\,24 bubble, G341.220-0.213, and G341.217-0.237 has velocities between --48.0 \kms\ and --40.0 \kms, compatible with a kinematical distance of 3.7 kpc previously adopted for the region. 

The gas distribution reveals a shell-like molecular structure of  $\sim$0.8 pc in radius detected in the velocity interval from --46 \kms\ to --41.5 \kms, encircling the S\,24 bubble. A cold dust counterpart of this shell is  detected  at wavelengths larger than 160 $\mu$m, i.e. in the LABOCA and SPIRE images. On the contrary, PACS emission at 70 $\mu$m, as well as MIPS emision at 24 $\mu$m from warm dust appear projected inside the bubble. Dust temperatures derived from the emission at 70 and 160  $\mu$m are in the range 26 to 60 K, whith the higher values closer to the center of the bubble. The detection of radio continuum emission  and the presence of warm dust indicates the existence of exciting sources and that the bubble is a compact \hii\ region. We estimated a molecular gas and $H_{\rm 2}$ ambient density in the shell of 3100 \msun\ and 2.5$\times$10$^4$ \cmtres. A search for excitation sources allows the identification of a massive star candidate projected onto the \hii\ region, which would provide the ultraviolet photons to heat the gas and dissociate and ionize the gas. A number of YSO candidates appear projected onto the molecular shell.

Part of the molecular gas bordering the S\,24 bubble coincides with the extended IRDC SDC341.194-0.221, for which we derived a kinetic temperature of about 20 K and a volume density of 10$^4$ \cmtres, based on LVG analysis. 

As regards G341.220-0.213, which resembles a horseshoe open towards the east, the  $^{13}$CO(3-2) line emission allowed to detect an arc-like  structure  with velocities in the range --43.8 \kms\ to --42.0 \kms, encircling the brightest sections of the IR source. The morphology indicates that the molecular gas is interacting with the IR source. A  dust counterpart is detected in the Herschel (SPIRE and PACS) and LABOCA images. The spectral energy distribution constructed using Herschel and ATLASGAL fluxes allows to derive a dust temperature of 28$\pm$1 K. A molecular mass of 1550 \msun\ is linked to this source, compatible with the mass derived from ATLASGAL. The H$_2$ ambient density amounts to 6.6$\times$10$^4$ \cmtres, in agreement with the ambient density value estimated from LVG analysis. A number of IR point sources having characteristics of YSO coincide with this region, which along with the presence of maser emission indicates that star formation is active. The SED for the derived fluxes suggests the presence of a massive central source of 10 \msun, which would be responsible for heating the dust and dissociating the molecular gas.

\gdos\ also coincides with molecular gas and  cold dust. The line emission allowed the detection of a molecular  clump  present at the interface between the S\,24 \hii\ region and G341.217-0.237, shapping the eastern border of the IR bubble. A high ambient density (4.2$\times$10$^5$ \cmtres) was estimated for the clump.
 
The total  mass of molecular gas in the region and the $H_2$ ambient density amount to 10300 \msun\ and 5900 \cmtres, indicating that \guno, \gdos, and the S\,24 \hii\ region are evolving in a high density region.

Finally, we investigated if a {\it collect and collapse} scenario can explain the presence of YSOs in the dense envelope around the S\,24 \hii\ region. We conclude that neither this mechanism nor the RDI process seem to occur in this region. Very probably, the \hii\ region is too young for triggering to have begun.
  
\acknowledgements 
We acknowledge the anonymous referee for his/her comments.  The ATLASGAL project is a collaboration between the Max-Planck-Gesellschaft, the European Southern Observatory (ESO) and the Universidad de Chile. C.E.C. acknowledges the kind hospitality of  M. Rubio and her family during her stay in Chile. L.C. acknowledges support from DIULS. V.F. acknowledges support from ESO-Chile Joint Committee and DIULS and  would like to thank Ivan Valtchanov, Bruno Altieri, and Luca Conversi for their support and valuable assistance in Herschel data processing. V.F. also acknowledges support from the Faculty of the European Space Astronomy Centre (ESAC).
This project was partially financed by CONICET of Argentina under project PIP 0356, UNLP under project 11/G120, and CONICyT of Chile through FONDECYT grant No. 1140839. 
This work is based [in part] on observations made with the Spitzer Space Telescope, which was operated by the Jet Propulsion Laboratory, California Institute of Technology  under a contract with NASA. This publication makes use of data products from the Two Micron All Sky Survey, which is a joint project of the University of Massa chusetts and the Infrared Processing and Analysis Center/California Institute of  Technology, funded by the National Aeronautics and Space Administration and the  National Science Foundation. The MSX mission was sponsored by the Ballistic Missile Defense Organization (BMDO).

\bibliographystyle{aa}
\bibliography{bibliografia-s24}
 
\IfFileExists{\jobname.bbl}{}
{\typeout{}
\typeout{****************************************************}
\typeout{****************************************************}
\typeout{** Please run "bibtex \jobname" to optain}
\typeout{** the bibliography and then re-run LaTeX}
\typeout{** twice to fix the references!}
\typeout{****************************************************}
\typeout{****************************************************}
\typeout{}

}

\end{document}